\documentclass[sn-apa]{sn-jnl}

\usepackage{graphicx}%
\usepackage{multirow}%
\usepackage{amsmath,amssymb,amsfonts}%
\usepackage{amsthm}%
\usepackage{mathrsfs}%
\usepackage[title]{appendix}%
\usepackage{xcolor}%
\usepackage{textcomp}%
\usepackage{manyfoot}%
\usepackage{booktabs}%
\usepackage{algorithm}%
\usepackage{algorithmicx}%
\usepackage{algpseudocode}%
\usepackage{listings}%
\usepackage{tikz}
\usepackage{comment}
\usepackage{bbm}
\usepackage{xspace}
\usepackage{float}

\newtheorem{prop}{Proposition}%
\newtheorem{lem}{Lemma}%
\newtheorem{corollary}{Corollary}%
\newtheorem{example}{Example}%
\newtheorem{definition}{Definition}%

\definecolor{mBlueDD}{RGB}{8,69,148}
\definecolor{mBlueDd}{RGB}{33,113,181}
\definecolor{mBlueD}{RGB}{66,146,198}
\definecolor{mBlued}{RGB}{107,174,214}
\definecolor{mBluel}{RGB}{158,202,225}
\definecolor{mBlueL}{RGB}{198,219,239}
\definecolor{mBlueLl}{RGB}{222,235,247}
\definecolor{mBlueLL}{RGB}{247,251,255}

\newcommand{\aggregationRule}{r}

\newcommand{\nbVoters}{v}
\newcommand{\voterName}{v}
\newcommand{\voterSet}{V}

\newcommand{\budgetingScenario}{E}
\newcommand{\nbProjects}{n}
\newcommand{\projectSet}{A}
\newcommand{\projectName}{a}
\newcommand{\costFunction}{c}
\newcommand{\cost}[1]{\costFunction(#1)}
\newcommand{\budgetLimit}{l}
\newcommand{\approvalSet}[1]{A_{#1}}
\newcommand{\bundle}{B}
\newcommand{\bundleIntersection}[1]{\bundle_{#1}}

\newcommand{\swapProjects}[2]{(\projectName_{#1} \leftrightarrow \projectName_{#2})}

\newcommand{\rate}[2]{r(#1,#2)}

\newcommand{\utilityFunction}{u}
\newcommand{\utilityFunctionMobius}{u_M}
\newcommand{\utility}[2]{\utilityFunction(#1,#2)}
\newcommand{\utilityScenario}{u^{\budgetingScenario}}
\newcommand{\utilityMobius}[2]{\utilityFunction_{M}(#1,#2)}

\newcommand*{\project}[4]{
	\draw (#3-#2,#4+0.1) rectangle (#3,#4+0.6);
	\draw (#3-#2/2,#4+0.35) node {#1};
}

\newcommand*{\projectColor}[5]{
	\draw[#5,line width=1.5pt] (#3-#2,0.1+#4) rectangle  (#3,0.6+#4);
	\draw (#3-#2/2,#4+0.35) node {\textcolor{#5}{#1}};
}

\begin{document}

\title[Project Interactions in Participatory Budgeting]{Detecting and taking Project Interactions into account in Participatory Budgeting}


\author*{\fnm{Martin} \sur{Durand}}\email{martin.durand@lip6.fr}

\author{\fnm{Fanny} \sur{Pascual}}\email{fanny.pascual@lip6.fr}

\affil{\orgdiv{Sorbonne Université}, \orgname{LIP6, CNRS}, \orgaddress{\street{5 Place Jussieu}, \city{Paris}, \postcode{75005}, \country{France}}}


\abstract{The aim of this paper is to introduce models and algorithms for the Participatory Budgeting problem when projects can interact with each other. 
In this problem, the objective is to select a set of projects that fits in a given budget. Voters express their preferences over the projects and the goal is then to find a consensus set of projects that does not exceed the budget. 
Our goal is to detect such interactions thanks to the preferences expressed by the voters.
Through the projects selected by the voters, we detect positive and negative interactions between the projects by identifying projects that are consistently chosen together. 
In presence of project interactions, it is preferable to select projects that interact positively rather than negatively, all other things being equal.  
We introduce desirable properties that utility functions should have in presence of project interactions and we build a utility function which fulfills the desirable properties introduced. 
We then give axiomatic properties of aggregation rules, and we study three classical aggregation rules: the maximization of the sum of the utilities, of the product of the utilities, or of the minimal utility. We show that in the three cases the problems solved by these rules are NP-hard, and we propose a branch and bound algorithm to solve them. We conclude the paper by experiments. }

\keywords{Computational social choice, Participatory Budgeting}



\maketitle

\newcommand{\alpharule}{$\alpha\!-\!\aggregationRule_{\utilityFunction}$}
\newcommand{\sumrule}{$\sum\!-\!\aggregationRule_{\utilityFunction}$}
\newcommand{\prodrule}{$\prod\!-\!\aggregationRule_{\utilityFunction}$}
\newcommand{\minrule}{$\min\!-\!\aggregationRule_{\utilityFunction}$}

\newcommand{\alphaproblem}{PB-{\sc Max}$\!-\!\alpha\!-\!{\utilityFunction}$\xspace}
\newcommand{\sumproblem}{PB-{\sc Max}$\!-\!\sum\!-\!{\utilityFunction}$\xspace}
\newcommand{\prodproblem}{PB-{\sc Max}$\!-\!\prod\!-\!{\utilityFunction}$\xspace}
\newcommand{\minproblem}{PB-{\sc Max}$\!-\!\min\!-\!{\utilityFunction}$\xspace}

\newcommand{\alphaproblemdec}{\alphaproblem-dec\xspace}
\newcommand{\sumproblemdec}{\sumproblem-dec\xspace}
\newcommand{\prodproblemdec}{\prodproblem-dec\xspace}
\newcommand{\minproblemdec}{\minproblem-dec\xspace}

\newcommand{\maxutilityproblem}{\textsc{PBSynergy} }
\newcommand{\maxutilityrule}{\textsc{PBSynergyRule} }
\newcommand{\cliqueMproblem}{\textsc{CliqueSparse} }

\newcommand{\setfeasiblesolutions}{\mathcal{S}_B^P}
\newcommand{\maximalcommonsubset}[2]{\mathcal{S}^*_{{#1,#2}}}

\section{Introduction}
\label{sec:pb_synergy_introduction}


Participatory budgeting is a democratic process in which community members decide how to spend part of a public budget. Started in Porto Alegre, Brazil, in 1989, this process has spread to over 7,000 cities around the world, and has been used to decide budgets from states, cities, housing authorities, universities, schools, and other institutions\footnote{
https://www.participatorybudgeting.org/}. The principle is  the following one: the authorities of a given community (e.g. a city, or a university) decide to dedicate a budget $\budgetLimit$ between  projects proposed by the community members. Some community members (e.g. citizens, or students)   propose projects, and write a proposal presenting their project and estimating its cost. All the community members are then asked to vote on the projects. There are several ways to collect voters' preferences. Due to its simplicity, the most widely used method is \emph{approval voting}, in which voters are asked to approve or not each of the proposed projects. A variant of this method, called \emph{knapsack voting}~\citep{goel2019knapsack}, and that we will consider in this paper, asks the voters to approve projects up to the budget limit~$\budgetLimit$:  
with knapsack voting, each voter is encouraged to give the set of projects that he or she would like to be selected, given the budget allocated. 
We start by reviewing existing work on participatory budgeting. Once the preferences of the voters have been expressed, the authorities use an algorithm which aggregates them and returns a set of projects (a bundle) of total cost at most~$\budgetLimit$. In practice, e.g. in Warsaw, the projects are usually  selected by decreasing number of votes. We start by reviewing existing work on participatory budgeting.

\subsection{Related work}

Participatory budgeting is a very active field in computational social choice and numerous other algorithms have been proposed~\citep{aziz2021participatory,aziz2017proportionally,peters2021proportional,talmon2019framework}. Several social welfare functions have been considered. The aim is usually either to   maximize the minimal utility of a voter~\citep{sreedurga2022maxmin}; to guarantee proportional representation to groups of voters with common interest~\citep{peters2021proportional,aziz2017proportionally,freeman2021},  both aiming to return ``fair" solutions;  
to maximize 
the sum of the utilities of the voters (utilitarian welfare);  or to maximize the products of these utilities (Nash product)~\citep{benade2021preference,goel2019knapsack,aziz2021participatory}. In this paper  we are interested in optimizing three of the most classical criteria: the maximization of the sum of the utilities, of the product of the utilities, or of the minimal utility of the voters. 

There are two main ways to define the utility of a voter. The first way defines the utility of a voter as the number of funded projects that he or she approves~\citep{peters2021proportional,jain2020participatory}. The second way defines the utility of a voter as the total amount of money allocated to projects approved by the voter~\citep{goel2019knapsack,talmon2019framework,freeman2021}. This second way of measuring the satisfaction of a voter is particularly relevant in the case of knapsack voting, where each voter can only approve a total budget of $\budgetLimit$: if a voter chooses to approve a project with a large cost at the expense of projects with smaller costs, it means that he or she prefers the large project to the smaller ones. We will consider this way to measure utilities. 


\smallskip

{\bf Project interactions.} 
Project interactions (also called synergies between projects) have been little explored so far.  In almost all the papers, it is assumed that the utility of a bundle (a set of projects) for a given voter is the sum of the utilities of these projects (number of projects or total cost of these projects, depending on the model considered). 
In a recent paper,  
\cite{fairstein2023participatory} 
   do an empirical study of several voting formats, without 
  considering   synergies. However, they say in their conclusion that ``real voter utilities likely exhibit complementarities and externalities — a far cry from our utility proxies''. 
Indeed, in practice, positive and negative synergies do exist. For example, two projects which are facilities that are planned to be built in the same location, or two projects which are very similar (e.g., two projects of playgrounds, or two skateboard parks) will have negative synergies: for a given voter, the utility of such two projects $A$ and $B$ will be smaller than the sum of the utilities of $A$ and $B$. On the contrary, some  projects are complementary and therefore have positive synergies. This is for example the case when a project aims to build a bicycle garage and another project aims to build a  meeting place nearby. For a given voter, the utility of two projects $A$ and $B$ with positive synergies will be larger than the sum of the utilities of $A$ and $B$. For two projects $A$ and $B$ which are independent, i.e., do not have positive neither negative synergies, the utility of the two projects $A$ and $B$ will be as usual the sum of the utilities of $A$ and $B$.

To the best of our knowledge, there are only two papers which deal with projects interactions~\citep{rey2023computational}. 
\cite{jain2020participatory} introduce a model in which they  
assume that the synergies between the projects are already known and are defined as a partition $P$ over the projects. The projects which belong to a same set of the partition  either have a substitution effect (i.e. a negative interaction) or a complimentary effect (a negative interaction). 
The authors define a utility function $f$ such that $f(i)$ is the utility that a voter $v$ gets from a set of the partition $P$ if $i$ projects from this set and approved by $v$ are in the returned bundle. 
If $f$ is concave (i.e. $f(i+1)-f(i)<f(i)-f(i-1)$) then projects in the same set of $P$ have negative interaction; if $f$ is convex (i.e. $f(i+1)-f(i)>f(i)-f(i-1)$)  then projects in the same set of $P$ have positive  interaction. 
The utility of a voter is the sum of the utilities it has over the different sets of the partition. This model is the first one to consider project interactions. In a subsequent paper, ~\cite{jain2021partition}, assuming such an existing
partition of the projects to interaction structures, take voter preferences to find such interaction structures (in their model, voters submit interaction structures, and the goal is to find an aggregated structure). 
~\cite{Fairsteinarxiv} also consider an underlying partition structure and ask the voters to give a partition of projects into groups of substitutes projects
: in this setting only negative interactions are considered. 

These papers are the first ones to consider and model project interaction. However, by partitioning the projects, their model  cannot represent situations in which a project $A$ can be both in positive interaction with a project $B$ and in negative interaction with a project $C$, situation that we wish to take into account in this paper. Furthermore, the authors of the previous mentioned papers 
assume that such a partition is either known~\citep{jain2020participatory}, or computed thanks to the partitions of the projects asked to the voters~\citep{jain2021partition,Fairsteinarxiv}, which can be a fastidious and complicated task for the voters.

\subsection{Our approach to interaction detection}
Our aim is not only to take into account interactions between projects into the utilities of voters, but also to detect the interactions through the preferences of the voters. Detecting such interactions through the votes is not possible if, as in ~\citep{jain2020participatory}, the voters use approval voting to give their opinions on the projects. Indeed, with approval voting, a voter tends to evaluate each project individually and to select the projects that he or she finds interesting according to his or her own criteria. Thus, it is likely that a voter who would like to see a playground built near his or her home will support all the playgrounds  projects, even if such projects interact negatively. On the contrary, with knapsack voting, each voter is asked how he or she would spend the money if he or she had the opportunity to decide. In that context, it is unlikely that a voter  selects projects that interact negatively, and on the contrary it is likely that projects that interact positively will be chosen. 
We think the best way to get reliable preferences (which express synergies) is to ask the following question to the voters: ``How would you spend the budget if you could make the decision ?''. Assuming most voters follow this recommendation, the synergies should be estimated quite accurately.

Detecting synergies can be done through the ballots approved by the voters, by looking at 
the frequencies of occurrence of groups of projects among the projects approved by the same voter, compared to the ``expected'' frequencies of this group of projects. If, for example, two projects $A$ and $B$ are selected together very often, we will deduce that they probably are in positive synergy. On the contrary, if two objects are never selected together, the synergy will be negative. Thus, by comparing the frequency of appearance of these projects $A, B$ together with the product of the frequency of $A$ and the frequency of $B$, we deduce synergies from the voters' choices. 

\medbreak

\begin{example}
Consider 
a budget $\budgetLimit\!=\!9$ and 
5 projects $\{A,\dots,E\}$ of costs $(2,3,3,1,1)$ (i.e. project $A$ has cost 2, while project $E$ has cost 1). 
Consider the following votes of $4$ voters: $\{A,B,D,E\}$, $\{A,B,C\}$,$\{C,E\}$,$\{A,B,D\}$. Each project has been selected 2 or 3 times but projects $A$ and $B$ are always selected together, and projects $C$ and $D$ are never selected in a same ballot: we will deduce that projects $A$ and $B$ have  a positive synergy while projects $C$ and $D$ have a negative synergy. Hence, whereas both bundles $\{A,B,C,E\}$ and $\{A,B,C,D\}$ are optimal for the utilitarian welfare, bundle $\{A,B,C,E\}$ is preferable 
because  $C$ and $D$ have a negative synergy while $C$ and $E$ do not.
\end{example}

\medbreak


One could argue that two projects will not be chosen by the same voter because of the budget limit and not because they have a negative interaction. 
First, if the sum of the costs of these two projects is larger than $\budgetLimit$, then these two projects will anyway not be chosen in the returned bundle. Second, we examined the costs distribution of projects from the $247$ real-world instances of knapsack voting from Pabulib~\citep{stolicki2020pabulib}. 
These instances mainly have 
``small projects": the vector of costs of projects of these instances is in average: $(0.56, 0.18, 0.09, 0.06, 0.04, 0.02, 0.02, 0.01, 0.01, \linebreak 0.01 )$ -- which means than $56\%$ of the projects have a cost between $0$ and $10\%$ of the budget, $18\%$ of the projects have a cost between $10$ and $20\%$ of the budget, and so forth. 
Additionally, on the same instances, the average (resp. median) total cost of the projects selected by a voter represents 66\% (resp. 75\%) of the budget. This means that a majority of voters could have selected one more project, and this among most of the unapproved  projects. Therefore, the overall low cost of the projects paired with the budget left unused in the votes suggests that if two projects are rarely selected together, it is usually not because of theirs costs.  
 \smallskip


Note that taking account of synergies between the projects  may be interesting even if all the projects have the same cost, as shown by the following example. 

\medbreak

\begin{example}
  Let us consider a scenario with 12 voters, $8$ projects of cost $1$ and a budget of $4$. Six voters select projects $1$ and $2$ plus a pair of projects in $\{5,6,7,8\}$, different for each one. The six other voters select projects $3$ and $4$ plus a pair in $\{5,6,7,8\}$, different for each one. Therefore, each project is selected exactly $6$ times. 
  
  \begin{figure}[H]
        \centering
        \begin{tikzpicture}
        \projectColor{$p_1$}{1}{1}{5}{mBlueD}
        \projectColor{$p_2$}{1}{2}{5}{mBlueD}
        \project{$p_5$}{1}{3}{5}
        \project{$p_7$}{1}{4}{5}

        \projectColor{$p_1$}{1}{1}{4}{mBlueD}
        \projectColor{$p_2$}{1}{2}{4}{mBlueD}
        \project{$p_5$}{1}{3}{4}
        \project{$p_8$}{1}{4}{4}

        \projectColor{$p_1$}{1}{1}{3}{mBlueD}
        \projectColor{$p_2$}{1}{2}{3}{mBlueD}
        \project{$p_6$}{1}{3}{3}
        \project{$p_7$}{1}{4}{3}

        \projectColor{$p_1$}{1}{1}{2}{mBlueD}
        \projectColor{$p_2$}{1}{2}{2}{mBlueD}
        \project{$p_6$}{1}{3}{2}
        \project{$p_8$}{1}{4}{2}

        \projectColor{$p_1$}{1}{1}{1}{mBlueD}
        \projectColor{$p_2$}{1}{2}{1}{mBlueD}
        \project{$p_7$}{1}{3}{1}
        \project{$p_8$}{1}{4}{1}

        \projectColor{$p_1$}{1}{1}{0}{mBlueD}
        \projectColor{$p_2$}{1}{2}{0}{mBlueD}
        \project{$p_5$}{1}{3}{0}
        \project{$p_6$}{1}{4}{0}

        \projectColor{$p_3$}{1}{6}{5}{mBlueD}
        \projectColor{$p_4$}{1}{7}{5}{mBlueD}
        \project{$p_5$}{1}{8}{5}
        \project{$p_7$}{1}{9}{5}

        \projectColor{$p_3$}{1}{6}{4}{mBlueD}
        \projectColor{$p_4$}{1}{7}{4}{mBlueD}
        \project{$p_5$}{1}{8}{4}
        \project{$p_8$}{1}{9}{4}

        \projectColor{$p_3$}{1}{6}{3}{mBlueD}
        \projectColor{$p_4$}{1}{7}{3}{mBlueD}
        \project{$p_6$}{1}{8}{3}
        \project{$p_7$}{1}{9}{3}

        \projectColor{$p_3$}{1}{6}{2}{mBlueD}
        \projectColor{$p_4$}{1}{7}{2}{mBlueD}
        \project{$p_6$}{1}{8}{2}
        \project{$p_8$}{1}{9}{2}

        \projectColor{$p_3$}{1}{6}{1}{mBlueD}
        \projectColor{$p_4$}{1}{7}{1}{mBlueD}
        \project{$p_7$}{1}{8}{1}
        \project{$p_8$}{1}{9}{1}

        \projectColor{$p_3$}{1}{6}{0}{mBlueD}
        \projectColor{$p_4$}{1}{7}{0}{mBlueD}
        \project{$p_5$}{1}{8}{0}
        \project{$p_6$}{1}{9}{0}
        \end{tikzpicture}
    \caption{Example with $\budgetLimit=4$}
    \end{figure}
    Without synergies, each bundle of $4$ projects is optimal for the sum of the utilities. However, using synergies, we can detect that $\{1,2\}$ and $\{3,4\}$ are probably two strong pairs in comparison to the others. We can also see that the subset$\{1,2,3,4\}$ is never chosen as a whole which may indicate an antisynergy of the complete subset.
\end{example}

\medbreak

In the sequel, we will sometimes consider the $k$-additivity hypothesis, which means that there are synergies between groups of up to $k$ projects. For example, with the $2$-additivity hypothesis, we consider only interactions between pairs of projects, and not between more important groups of projects. In addition to the fact that it is realistic  
that synergies are important only for small values of $k$, considering this hypothesis will have repercussions on the complexity of our algorithms.  

We conclude this introduction with an example showing that, in practice, positive (resp. negative) interactions may indeed be detected through the frequencies of co-occurrence of the projects in the same bundles. 

\medbreak

\begin{example}
By looking at real-world knapsack voting instances in the Pabulib  library~\citep{stolicki2020pabulib}, and by considering that two projects interact positively (resp. negatively) when they are (resp. are not) chosen together, we identified several cases in which projects seem to interact positively or negatively\footnote{To be precise, to detect these interactions, we used the utility function $\utilityFunctionMobius$ presented in Section 4, by considering $2$-additivity hypothesis.}. For example, in Warsaw (poland\_warszawa\_2017\_niskie-okecie.pb), two projects for the same neighbourhood, the first one being building a sport court and the second one building a playground, were chosen together less often than expected (given how often each one was individually approved). Our model says  that they interact negatively, which makes sense, these projects being close to being susbtitutes. In another instance (poland\_warszawa\_2018\_niskie-okecie.pb), two projects, the first one being building alleys in a park and the second one building public lightning in the same park, were consistently chosen together, which our model interpreted as a positive synergy. This also makes sense since these projects are clearly complementary. 
\end{example}

\subsection{Overview of our results}
We tackle the indivisible participatory budgeting problem, with knapsack voting, by considering that projects are not independent, but that there may have positive and negative synergies between them.

\begin{itemize}
    \item In Section~\ref{sec:utility_function}, we  propose  desirable properties for utility functions in presence of project interactions. 
    \item In Section~\ref{sec:utility_example} we present a particular utility function, derived from Möbius transforms and denoted by $\utilityFunctionMobius$, that fulfills the axioms defined on the previous section.
    \item In Section~\ref{sec:pb_methods_axioms}, we study axiomatic properties of aggregation rules.  We consider in particular three aggregation rules, which either maximize the sum of the utilities, the product of the utilities, or the minimal utility of the voters. 
    \item In Section~\ref{sec:complexity} 
    we show that these rules solve NP-hard problems, and that synergies make the problem harder since it is NP-hard to maximize the sum of the utilities with unit size projects when there are synergies, whereas this problem can be solved easily  without synergies. These results hold for utility function $\utilityFunctionMobius$ but also for other very general synergy functions.
    \item In Section~\ref{sec:branch_and_bound}, we propose an exact branch and bound algorithm which can be used with any utility function, and we conclude with an experimental evaluation.
\end{itemize}

\section{Preliminaries}
\label{sec:pb_synergy_preliminaries}
We use the general framework for approval-based participatory budgeting proposed by ~\cite{talmon2019framework}. 
A \emph{budgeting scenario} is a tuple $\budgetingScenario\!=\!(\projectSet,\voterSet,\costFunction,\budgetLimit)$ where $\projectSet=\{\projectName_1, \dots , \projectName_{\nbProjects}\}$ is a set of $\nbProjects$ \emph{projects}, or items, and $\costFunction: \projectSet \rightarrow \mathbbm{N}$ is a \emph{cost function}:   $\cost{\projectName}$ is the cost of project $\projectName \in \projectSet$  -- abusing the notation, given a subset $S$, we denote by $\cost{S}$ the total cost of $S$: $\cost{S}\!=\!\sum_{\projectName \in S} \cost{a}$. The budget limit  is $\budgetLimit \in \mathbbm{N}$. The set $\voterSet\!=\!\{\voterName_1, \dots , \voterName_{\nbVoters}\}$ is a set of $\nbVoters$ voters. Each voter $\voterName_i \in \voterSet$ gives a set of approved projects $\approvalSet{i} \subseteq \projectSet$, containing a set of projects that she approves of and such that $\cost{\approvalSet{i}} \leq \budgetLimit$. We denote by $\mathcal{E}^{\projectSet}$ the set of all possible budgeting scenarios having $\projectSet$ as a set of projects.

A \emph{budgeting method} $\aggregationRule$ is a function taking a budgeting scenario $\budgetingScenario\!=\!(\projectSet,\voterSet,\costFunction,\budgetLimit)$ and returning a \emph{bundle} $\bundle \subseteq \projectSet$ such that 
$\cost{\bundle} \leq \budgetLimit$. 
We consider that a budgeting method always returns a unique bundle (we can use usual tie-breaking techniques to handle instances with several winning bundles).
The winning bundle for a budgeting scenario $\budgetingScenario$ is denoted by $\aggregationRule(\budgetingScenario)$. A project is \emph{funded} if it is contained in the winning bundle $\bundle$. Given a bundle $\bundle$ and a voter $\voterName_i$ with her approval set $\approvalSet{i}$, we denote by  $\bundleIntersection{i}\!=\!\approvalSet{i} \cap \bundle$ the set of  projects common to $\approvalSet{i}$ and $\bundle$.
\smallskip
  {\bf Utility functions. }
A \emph{utility function} $\utilityFunction: 2^{\projectSet} \rightarrow \mathbbm{R}_0^+$ is a set function which gives a value to each  subset of items. 
A \emph{linear utility function} is such that that the value of a bundle $\bundle$ is the sum of the utilities of its items: $\utilityFunction(\bundle)\!=\!\sum_{\projectName \in \bundle} \utilityFunction(\{\projectName\})$. 
The \emph{overlap utility} function, introduced for the knapsack voting by \cite{goel2019knapsack}, 
considers that the utility of a bundle $\bundle$ is the sum of the  costs of the projects in $\bundle$: $f(\approvalSet{i},\bundle)\!=\!\sum_{\projectName \in \bundleIntersection{i}}  \cost{\projectName}$
 

  \smallskip
  {\bf Satisfaction functions. }
A \emph{satisfaction function} $f$ is a function $f:2^{\projectSet} \times 2^{\projectSet} \rightarrow \mathbbm{R}$, which, given a voter $\voterName_i \in \voterSet$ and a bundle $\bundle \subseteq \projectSet$, returns  
the satisfaction that $\voterName_i$ gets from $\bundle$. 
Given a selected bundle $\bundle$ and a utility function $\utilityFunction$, we will consider that the satisfaction of voter $\voterName_i$ from bundle $\bundle$ is the utility of $\approvalSet{i} \cap \bundle$:  $f(\approvalSet{i},\bundle)\!=\!\utilityFunction{\bundleIntersection{i}}$.

. 

The utility function aims at associating to each possible bundle an evaluation of its quality. The satisfaction function indicates, given two sets of projects, the first one being the preference of a voter and the other being a potential solution, how satisfied the voter is given the solution.

In the sequel, we will consider 
generalizations of the overlap utility function that take into account potential projects interactions. Since these function may depend on the instance, we will denote the utility of the subset $\bundleIntersection{i}$ as: $\utility{\bundleIntersection{i}}{\budgetingScenario}$.

 \smallskip
  {\bf Aggregating criterion. }In order to obtain a solution satisfying the whole population, we study the three most classical  aggregating methods: the sum ($\sum$), the product ($\prod$) and the minimum ($\min$) of the satisfactions of the voters. We 
 denote by \alpharule\xspace  the budgeting method returning the $\alpha$ aggregation of the utility function $\utilityFunction$, where $\alpha \in \{\sum, \prod, \min\}$. This rule returns an optimal bundle of the associated maximization optimization problem, that we will call problem \alphaproblem\space (e.g. problem PB-{\sc Max}$\!-\!\sum\!-\!u$ consists in computing a bundle maximizing the sum of the utilities of the voters when the utility function used is $\utilityFunction$).
These three aggregating concepts rely on different ideas of the collective satisfaction. The sum criterion maximizes the average satisfaction of a voter. 
The minimum tries to satisfy as much as possible the least satisfied voter -- this is an egalitarian view. Finally the product stands in between the two previous criteria: the product is very penalized by the presence of very low utility values, however, it still takes into account the larger values. This last criterion has been the favourite of the voters in an experimental study~\citep{experiment2021}. These three criteria share several axiomatic and computational properties, as we will see in the following sections.

We will now discuss how to obtain satisfactory utility functions and how mathematical properties on such functions impact the budgeting methods.

\section{Axioms for utility functions}
\label{sec:utility_function}

In this section, we define desirable properties for utility functions in the presence of synergies. 

The first property states that the utility of a single project should be proportional to its cost. This property is fulfilled by the \emph{overlap utility} function~\citep{goel2019knapsack}. 
It is particularly meaningful in knapsack voting: 
since there is a budget constraint on the approval set of the voters, the approval of a project is done with full knowledge of its cost and the approval of a costly project is done at the expense of the budget for other projects. 

\medbreak

\begin{definition}
Given a budgeting scenario $\budgetingScenario\!=\!(\projectSet,\voterSet,\costFunction,\budgetLimit)$, a utility function $\utilityScenario: 2^{\projectSet} \times \mathcal{E}^A 
\rightarrow \mathbbm{R}^+_0$ is \emph{cost consistent} if there exists a constant $k$ such that for each project $\projectName$ in $\projectSet$, we have $\utility{\{\projectName\}}\!=\!k \cdot \cost{\projectName}$.
\end{definition}

The factor $k$ allows normalization. 
This property insures that the utility function follows the cost function for the sets containing only one project. 

The following classical property ensures that the utility of a set does not decrease when the set grows.
This ensures that we cannot decrease a voter satisfaction by adding a project that she selected.


\medbreak

\begin{definition}
Given a budgeting scenario $\budgetingScenario\!=(\projectSet,\voterSet,\costFunction,\budgetLimit)$, a utility function $\utilityScenario$ is \emph{super-set monotone} if for any subset $X_{sub}$ and $X$ such that $X_{sub}\subset X$, we have $\utility{X_{sub}}{\budgetingScenario}\leq \utility{X}{\budgetingScenario}$. 
\end{definition}

Relaxing the \emph{neutrality} principle~\citep{brandt2016handbook}, the next property states that two similar projects should be treated equally. Given a set $S$, we denote by $S_{\swapProjects{i}{j}}$ the set obtained from $S$ by swapping $\projectName_i$ and $\projectName_j$: $\projectName_i$ (resp. $\projectName_j$) belongs to $S_{\swapProjects{i}{j}}$ if and only $\projectName_j$ (resp. $\projectName_i$) belongs to $S$, and each project $\projectName_k \notin\{\projectName_i,\projectName_j\}$ belongs to $S_{\swapProjects{i}{j}}$ if and only if $\projectName_k$ belongs to $S$. 
We also denote by $\budgetingScenario_{\swapProjects{i}{j}}$ 
the budgeting scenario obtained from $\budgetingScenario$ by swapping the approval of the projects $\projectName_i$ and $\projectName_j$: a voter $\voterName_l$ approves $\projectName_i$ (resp. $\projectName_j$) in $\budgetingScenario^{\swapProjects{i}{j}}$ if and only if $\voterName_l$ approves $\projectName_j$ (resp. $\projectName_i$) in $\budgetingScenario$. 

\medbreak

\begin{definition}
Given a budgeting scenario $\budgetingScenario=(\projectSet,\voterSet,\costFunction,\budgetLimit)$, and two projects $\projectName_i$ and $\projectName_j$ of $\projectSet$ such that $\cost{\projectName_i}\!=\!\cost{\projectName_j}$,   a utility function $\utilityScenario$ is \emph{cost-aware neutral} if $\utility{S}{\budgetingScenario}\!=\!\utility{S_{\swapProjects{i}{j}}}{\budgetingScenario_{\swapProjects{i}{j}}}$.
\end{definition}

Note that this property is inspired by the \emph{Processing Time Aware neutrality} property used in the collective schedules model~\citep{durand2022collective}: this property  ensures that two tasks of equal processing time are treated equally. 
We restrict our analysis to cost-aware neutral utility functions since no pair of projects with the same cost should be treated differently.

If a subset of item is consistently chosen as a whole, then the utility it brings should be higher than the sum of the utilities of the items. On the opposite side, if projects are never chosen together, then the utility of the whole subset should be lower than the sum of utilities of the items. The third axiom states that the more a subset appear together, the more its utility should increase, everything else being equal.\\


The next property, the \emph{effect of positive synergies} ensures that the utility of subsets of projects that always appear together is larger than the sum of the utilities of its components.


\medbreak

\begin{definition}
Given a budgeting scenario $\budgetingScenario=(\projectSet,\voterSet,\costFunction,\budgetLimit)$, a utility function $\utilityScenario$ fulfills the \emph{effect of positive synergies} (resp. \emph{strong effect of positive synergies}) property if, for each subset $S$ in $2^{\projectSet}$ such that  for each voter $\voterName_i$ we have either $S \subseteq \approvalSet{i}$ or $S \cap \approvalSet{i}\!=\!\emptyset$ and such that there exists $\voterName_k \in \voterSet$ with $S \subseteq \approvalSet{k}$, then  $\utility{S}{\budgetingScenario} \geq \sum_{\projectName \in S} \utility{\{\projectName\}}{\budgetingScenario}$ (resp. $\utility{S}{\budgetingScenario} > \sum_{\projectName \in S} \utility{\{\projectName\}}{\budgetingScenario}$).
\end{definition}

The next property ensures that the utility of subsets of projects that never appear together  is smaller than (or equal to) the sum of the utilities of its components.

\medbreak

\begin{definition}
Given a budgeting scenario $\budgetingScenario=(\projectSet,\voterSet,\costFunction,\budgetLimit)$, a utility function $\utilityScenario$ fulfills the \emph{effect of negative synergies} (resp. \emph{strong effect of negative synergies}) property if, for each subset $S$ in $2^{\projectSet}$ such that for each  voter $\voterName_i \in \voterSet$ we have $|S \cap \approvalSet{i}|\leq 1$, then  $\utility{S}{\budgetingScenario} \leq \sum_{\projectName \in S} \utility{\{\projectName\}}{\budgetingScenario}$ (resp. $\utility{S}{\budgetingScenario} < \sum_{\projectName \in S} \utility{\{\projectName\}}{\budgetingScenario}$).
\end{definition}


The next property 
states that the utility of a subset should increase with the number of appearances of the whole subset in the preferences of voters with respect to a solution in which the number of approvals of the items is the same but the items are not approved by the same voters. 


\medbreak

\begin{definition}
Let $\budgetingScenario=(\projectSet,\voterSet,\costFunction,\budgetLimit)$ be a budgeting scenario, $S \subseteq \projectSet$ be a subset such that $\cost{S} \leq \budgetLimit$, and let $\voterName_i$ and $\voterName_j$ be two voters of $\voterSet$ such that $S \subseteq (\approvalSet{i} \cup \approvalSet{j})$, $\approvalSet{i} \cap \approvalSet{j}\!=\!\emptyset$, $S \not\subseteq \approvalSet{i}$, $S \not\subseteq \approvalSet{j}$,  and $\cost{\approvalSet{i} \cup \approvalSet{j} \setminus S} \leq \budgetLimit$. 
Let $\voterSet_S\!=\! \voterSet \cup \{\voterName_k, \voterName_l\} \setminus \{\voterName_i,\voterName_j\}$, where  $\voterName_k$ and $\voterName_l$ are two voters who are not in $\voterSet$ and such that $\approvalSet{k}\!=\!S$ and $\approvalSet{l}\!=\!(\approvalSet{i} \cup \approvalSet{j}) \setminus S$. Let $\budgetingScenario'\!=\!(\projectSet,\voterSet_S,\costFunction,\budgetLimit)$ be a budgeting scenario. A utility function $\utilityScenario$ satisfies \emph{regrouping monotonicity} if $\utility{S}{\budgetingScenario} < \utility{S}{\budgetingScenario'}$.
\end{definition}

We can also imagine creating  utility functions thanks to prior knowledge on the projects, however in such cases, it is possible that the last three properties are violated.

In the following section, we propose a utility function 
taking synergies into account, and that fulfills the properties that we have introduced in this section.


\section{A utility function taking synergies into account}
\label{sec:utility_example}

\subsection{A function using  Möbius transforms: $\utilityFunctionMobius$}

Möbius transforms~\citep{rota1964foundations} are a classical tool for measuring synergies in sets of items.  
Given a utility function $\utilityFunction: 2^{\projectSet} \rightarrow \mathbbm{R}_0^+$,  the Möbius transform of a subset $S$, denoted by $m(S)$, expresses the level of synergy between the items in $S$. 
For a set $S\!=\!\{a,b\}$ of two elements, and if $\utilityFunction(\emptyset)\!=\!0$, we have $m(S)\!=\!\utilityFunction(\{a,b\})\!-\!\utilityFunction(\{a\})\!-\!\utilityFunction(\{b\})$.  
More generally, the Möbius transform of a set $S$ is calculated as follows:
$$
m(S)\!=\!\sum_{C \subseteq S} (-1)^{|S \setminus \{C\}|} \utilityFunction(C)
$$

The Möbius transform $m(S)$ expresses the level of synergy between the elements of the subset $S$. 
If it is negative, this indicates a negative interaction between the elements of $S$; if it is null, this indicates independence of the elements; and if it is positive, this indicates positive interaction between the elements. 

\medbreak

\begin{example}
Let us consider a utility function $\utilityFunction$ over a set of items $\{1,2,3\}$. The utilities are as follows:
\begin{table}
    \centering
    \begin{tabular}{c|c|c|c|c|c|c|c|c}
         C & $\emptyset$ & $\{1\}$ & $\{2\}$ & $\{3\}$ & $\{1,2\}$ & $\{1,3\}$ & $\{2,3\}$ & $\{1,2,3\}$ \\
         \hline
         $\utilityFunction (C)$ & 0 & 0.2 & 0.4 & 0.5 & 0.5 & 0.7 & 0.8 & 1\\
    \end{tabular}
    \label{tab:ex_mobius}
\end{table}
Let us compute the Möbius transform of subset \{1,2\}
$$
\begin{array}{rclcccccc}
     m(\{1,2\})\!&\!\!=\! \!&\!(-1)^{0} u(\{1,2\})\!&\!+\!&\!(-1) u(\{1\})\!&\!+\!&\!(-1) u(\{2\})\!&\!+\!&\!(-1)^{2} u(\emptyset)  \\
     m(\{1,2\})\!&\!\!=\! \!&\!\hspace{1em}u(\{1,2\})\!&\!-\!&\! u(\{1\})\!&\!-\!&\!u(\{2\})\!&\!+\!&\!u(\emptyset)  \\
     
     m(\{1,2\})\!&\!=\!&\!-0.1 & & & & & & \\
\end{array}
$$

We find a negative Möbius transform, indicating a negative interaction between elements $1$ and $2$. 
\end{example}

{\bf A utility function from Möbius transforms.} 
It is not only possible to find the Möbius transforms from a utility function, it is also possible to build a utility function from the Möbius transforms thanks to the following expression~\citep{rota1964foundations}:
$$
\utilityFunction(S)\!=\!\sum_{C \subseteq S} m(C)
$$

The utility of a subset $S$ is then the sum of Möbius transforms of its elements -- which is also the sum of their utilities -- plus the Möbius transforms of the subsets included in $S$, representing their level of positive and negative synergies. 
Therefore, if we can measure the level of synergy of each subset, we can build a utility function. 

\smallskip
We use a statistical approach in order to infer synergies from the preferences. Let $\rate{S}{\voterSet}$ be  the rate of occurrence of a subset $S$ in the approval sets of voters in $\voterSet$ (i.e. the ratio between the number of voters who selected all the projects of $S$, and the total number of voters). The expected rate of occurrence of a whole subset $S$ if all of its elements were perfectly independent (ignoring possible cost constraints), would be $\prod_{\projectName \in S} \rate{\{\projectName\}}{\voterSet}$, the product of the appearance rates of each the elements of $S$. 
We use $(\rate{S}{\voterSet}-\prod_{\projectName \in S} \rate{S}{\voterSet})$ 
as a marker of synergy. If it is null, then the projects appear as independent in the preferences. If it is positive, then the subset appears more frequently than expected if the preferences were random, indicating a positive interaction. If it is negative, it indicates on the contrary a negative interaction.  

We set $\utilityFunction(\emptyset)\!=\!m(\emptyset)\!=\!0$ and, to insure cost consistency, we set  $\utilityFunction(\{a\})\!=\!m(\{a\})\!=\!\cost{\projectName}$. 
 Since  $(\rate{S}{\voterSet}-\prod_{\projectName \in S} \rate{S}{\voterSet})$ has a range included in $[-1;1]$, 
we  multiply this difference by the cost of the subset. We obtain: $m(S,\budgetingScenario)\!=\!(\rate{S}{\voterSet}-\prod_{\projectName \in S} \rate{S}{\voterSet}) \cdot \cost{S}$. 
We finally adapt  this definition so that the utility function obtained from the Möbius transforms fulfills super-set monotonicity:
\begin{equation}
    m(S,\budgetingScenario)\!=\!\left\{
    \arraycolsep=0.5pt
    \begin{array}{lll}
        0 & & \text{if } S\!=\!\emptyset \\
        \cost{\projectName} & & \text{if } S\!=\!\{\projectName\}\\
        \max\lbrace & (\rate{S}{\voterSet}-\prod\limits_{\projectName \in S} \rate{S}{\voterSet})\cost{S}, & \\
        & \max\limits_{\projectName \in S}\left(\!-\!\sum\limits_{C \subset S, \projectName \in C} m(C,\budgetingScenario)\right)\rbrace & \text{otherwise}
    \end{array}
    \right.
    \label{eq:mobius_transform}
\end{equation}

The intuition is the following one: $\sum\limits_{C \subset S, \projectName \in C} m(C,\budgetingScenario)$ is the sum of the Möbius transforms of subsets containing project $\projectName$. By ensuring that the Möbius transform of $S$ is larger than or equal to the opposite of this sum, we ensure that 
%
the utility of $S$ is not smaller than the utility of $S \setminus \{a\}$. 

Note that guaranteeing super-set monotonicity implies that we know the Möbius transforms value of smaller sets.  
The utility function $\utilityFunctionMobius$ is as follows:
\begin{equation}
    \utilityMobius{S}{\budgetingScenario}\!=\!\sum_{C \subseteq S} m(C,\budgetingScenario)
    \label{eq:utility_mobius}
\end{equation}

\subsection{Properties of $\utilityFunctionMobius$, and remarks on its computation} 
We use Equation~\ref{eq:utility_mobius} to determine the utility of a bundle with function $\utilityFunctionMobius$. Because of its recursive nature, we compute first, as a preprocessing step, the utility of singletons, then pairs, then triplets and so forth. Determining the utilities in this way costs up to $2^{\nbProjects}$ (since there are $2^{\nbProjects}$ subsets) times $\nbVoters \times \nbProjects$ operations (since determining the appearance rate of a subset costs $\nbVoters \times \nbProjects$ operations). This calculation is much faster with the $k$-additivity hypothesis, since the Möbius transform associated to any subset of size larger than $k$ is then $0$. 
Therefore, with such an hypothesis, 
we simply need to know the  Möbius transforms 
of the subsets of size at most $k$: the preprocessing part is polynomial if $k$ is a constant.

\nopagebreak 

We now state that the utility function $\utilityFunctionMobius$ fulfills all the desirable properties stated in Section 3. This is true even with the $k$-additivity assumption, for any value of $k$. 


\medbreak

\begin{prop}
The utility function $\utilityFunctionMobius$ fulfills \emph{cost consistency}, \emph{super-set monotonicity}, the \emph{effect of positive synergies} property,  the \emph{effect of negative synergies} property, \emph{regrouping monotonicity} and \emph{cost aware neutrality}. 
It also fulfills the strong effect of positive synergies property if for each project $\projectName$, there is at least one voter who does not select $\projectName$. 
\label{prop:utility_function_um}
\end{prop}

\begin{proof}

\begin{itemize}
    \item \textbf{Cost consistency.} As defined in equation~\ref{eq:mobius_transform}, the Möbius transform of a single project is its cost. Since the utility of a single project is its Möbius transform, assuming the Möbius transform and the utility of the empty set is $0$, the utility $\utilityMobius{\{\projectName\}}{\budgetingScenario}\!=\!\cost{\projectName}$ for any project $\projectName$. Therefore $\utilityFunctionMobius$ fulfills cost consistency.
    \item \textbf{Super-set monotonicity.} As detailed earlier, the super-set monotonicity of the function $\utilityFunctionMobius$ is insured by the definition of the Möbius transform. As a reminder, to fulfill super-set monotonicity, the function $\utilityFunctionMobius$ has to verify the following property: $\utilityMobius{S}{\budgetingScenario} \geq \utilityMobius{S \setminus \{\projectName\}}{\budgetingScenario}$ for all $S \in 2^{\projectSet} \setminus \emptyset$ and all $\projectName \in S$. Since, by definition, we have $m(S,\budgetingScenario) \geq - \sum_{C \subset S, \projectName \in C}m(C,E)$ for all $\projectName \in S$, it means that $\sum_{C \subseteq S, \projectName \in C} m(C,\budgetingScenario) \geq 0$ for all $\projectName \in S$ and therefore $\sum_{C \subseteq S} m(C,\budgetingScenario) \geq \sum_{C' \subseteq S \setminus \{\projectName\}} m(C',E)$ for all $\projectName \in S$. By definition of $\utilityFunctionMobius$, this means $\utilityMobius{S}{\budgetingScenario}\geq \utilityMobius{S\setminus \{\projectName\}}{\budgetingScenario}$ for all $\projectName \in S$. The utility function $\utilityFunctionMobius$ fulfills super-set monotonicity.
    \item \textbf{Effect of positive synergies.} Let $S$ be a subset of projects such that for any $\projectName \in S$ and any $\voterName_i \in \voterSet$, $\projectName \in \approvalSet{i} \implies S \subseteq \approvalSet{i}$ and such that $\exists \voterName_k \in \voterSet$ with $\projectName \in \approvalSet{k}$. In other words, if a voter approves of one of the elements of $S$, she approves of all projects in $S$ and such a voter exists in $\voterSet$. For such a subset, the value $\rate{S}{\voterSet}-\prod_{\projectName \in S} \rate{\{\projectName\}}{\voterSet}$ is equal to $\rate{S}{\voterSet}- \rate{S}{\voterSet}^{|S|}$. Since the $\rate{S}{\voterSet}$ value is larger than $0$ and smaller than or equal to $1$, the difference $\rate{S}{\voterSet}-\rate{S}{\voterSet}^{|S|}$ is positive or null. This means that $(\rate{S}{\voterSet}-\prod_{\projectName \in S} \rate{\{\projectName\}}{\voterSet})\cost{S}$ is positive or null, this means that the Möbius transform of $S$ is positive or null, $m(S,\budgetingScenario) \geq 0$. The same remark can be said about all subset $C \subseteq S$ since all the projects of $S$ are only selected together. Therefore, we have $\sum_{C \subseteq S, |C| \geq 2} m(C) \geq 0$. By definition, the Möbius transforms of the single projects are their cost, we then have: $\sum_{C \subseteq S} m(C) \geq \sum_{a \in S} \cost{\projectName}$, and consequently: $\utilityMobius{S}{\budgetingScenario} \geq 0$. The utility function $\utilityFunctionMobius$ fulfills the effect of positive synergies property. If we suppose that for each project $\projectName$, there is at least one voter who does not select $\projectName$, then for each subset $S$, there is at least one voter who does not select $S$. Then $\rate{S}{\voterSet}$ is smaller than $1$ and the difference $\rate{S}{\voterSet}-\rate{S}{\voterSet}^{|S|}$ is strictly positive. In this case, $\utilityFunctionMobius$ fulfills the strong effect of positive synergies property.
    \item \textbf{Effect of negative synergies.} Let $S$ be a subset of projects such that for any $\projectName \in S$ and any $\voterName_i \in \voterSet$, $\projectName \in \approvalSet{i} \implies S \cap \approvalSet{i}\!=\!\{\projectName\}$. In other words, if a voter selects an element $\projectName$ of $S$, then it is the only element of $S$ she selects. For such a subset, the value $\rate{S}{\voterSet}-\prod_{\projectName \in S} \rate{\{\projectName\}}{\voterSet}$ is negative or null, since $S$ never appears but the element of $S$ can appear individually. This is true for any subset $C \subseteq S$ such that $|C|\geq 2$. When summing the Möbius transforms all these subsets included in $S$,we will have the Möbius transforms of singleton that are positive and equal to the cost the project and then null or negative values. This means that the overall utility of $S$ cannot be greater than the sum of the utility of its components. Therefore, $\utilityFunctionMobius$ fulfills the effect of positive synergies property.
    \item \textbf{Regrouping monotonicity.} Let $\budgetingScenario=(\projectSet,\voterSet,\costFunction,\budgetLimit)$ be a budgeting scenario and let $S \in 2^{\projectSet}$ be a subset of projects with $\cost{S} \leq \budgetLimit$. Let $\voterName_i$ and $\voterName_j$ be two voters in $\voterSet$ such that $\approvalSet{i} \cap \approvalSet{j}\!=\!\emptyset$, $S \subseteq \approvalSet{i} \cup \approvalSet{j}$ and $\cost{\approvalSet{i} \cup \approvalSet{j} \setminus S} \leq \budgetLimit$. We consider voters $\voterName_k$ and $\voterName_l$ with $\approvalSet{k}\!=\!S$ and $\approvalSet{l}\!=\!\approvalSet{i} \cup \approvalSet{j} \setminus S$, and a set of preferences $\voterSet_S\!=\!\voterSet \cup \{\voterName_k,\voterName_l\} \setminus \{\voterName_i,\voterName_j\}$. Let $\budgetingScenario'\!=\!(\projectSet,\voterSet_S,\costFunction,\budgetLimit)$ be a budgeting scenario. In $\voterSet_S$ any subset $C \subseteq S$ appears at least as often than in $\voterSet$ and any project appears as much in $\voterSet_S$ than in $\voterSet$, therefore for any $C \subseteq S$, $\rate{C}{\voterSet_S} \geq \rate{C}{\voterSet}$, consequently $m(C,E) \geq m(C,E)$ and $u_M(C,E') \geq \utilityMobius{C}{\budgetingScenario}$. Since $u_M(C,E') \geq \utilityMobius{C}{\budgetingScenario}$, for all $C \subseteq S$ and $r(S,V_S) > \rate{S}{\voterSet}$ we see that $m(S,E') > m(S,\budgetingScenario)$ and therefore $\utilityMobius{S,\budgetingScenario'} > \utilityMobius{S}{\budgetingScenario}$ from equation~\ref{eq:utility_mobius}. Thus, $\utilityFunctionMobius$ fulfills regrouping monotonicity.
    \item \textbf{Cost aware neutrality.} 
    Given a budgeting scenario $\budgetingScenario=(\projectSet,\voterSet,\costFunction,\budgetLimit)$, let $\budgetingScenario_{\swapProjects{i}{j}}\!=\!(\projectSet,\voterSet_{\swapProjects{i}{j}},\costFunction,\budgetLimit)$ be the budgeting scenario obtained from $\budgetingScenario$ by swapping the approval of two projects $\projectName_i$ and $\projectName_j$ such that $\cost{\projectName_i}\!=\!\cost{\projectName_j}$. For a given subset $S$, let $S_{\swapProjects{i}{j}}$ be the subset obtained from $S$ by swapping $\projectName_i$ and $\projectName_j$, i.e. $S_{\swapProjects{i}{j}}$ contains the same projects than $S$ except for $\projectName_i$ and $\projectName_j$, if $S$ contains $\projectName_i$, $S^{\swapProjects{i}{j}}$ contains $\projectName_j$ and if $S$ contains $\projectName_j$, $S^{\swapProjects{i}{j}}$ contains $\projectName_i$. 
    Since $\cost{\projectName_i}\!=\!\cost{\projectName_j}$, and since for any subset $C$, $\rate{C}{\voterSet}\!=\!\rate{C_{\swapProjects{i}{j}}}{V_{\swapProjects{i}{j}}}$, we can see that $m(C,\budgetingScenario)\!=\!m(C_{\swapProjects{i}{j}},\budgetingScenario_{\swapProjects{i}{j}})$ and, because of equation~\ref{eq:utility_mobius} that $\utilityMobius{C}{\budgetingScenario}\!=\!\utilityMobius{C_{\swapProjects{i}{j}}}{E_{\swapProjects{i}{j}}}$. The utility function $\utilityFunctionMobius$ fulfills cost aware neutrality.

\end{itemize}
\end{proof}


\section{Axioms for budgeting methods}
\label{sec:pb_methods_axioms}

In this section we discuss some axiomatic properties of the different aggregation rules, relying on the properties of the utility function used. We try, when it is possible, to have general results relying on the properties introduced in section~\ref{sec:utility_function} instead of on specific utility functions. 
We start with the \emph{inclusion maximality} axiom \citep{talmon2019framework}, also known as \emph{exhaustiveness} \citep{aziz2017proportionally}. This axiom states that if a bundle $\bundle$ is a winning bundle according to a budgeting method $\aggregationRule$, then it is either exhaustive, in the sense that it is impossible to add a project without exceeding the budget limit, or 
any of its feasible superset is also a winning bundle.

\medbreak

\begin{definition}
A budgeting method $\mathcal{R}$ satisfies \emph{inclusion maximality} if for any budgeting scenario $\budgetingScenario=(\projectSet,\voterSet,\costFunction,\budgetLimit)$ and each pair of feasible bundles $\bundle$ and $\bundle'$ such that 
$B' \subset B$, it holds that $B' \in \mathcal{R}(E) \implies B \in \mathcal{R}(E)$.
\end{definition} 

\medbreak

\begin{prop}
If a utility function $\utilityFunction$ fulfills super-set monotonicity, then the budgeting method \alpharule, for $\alpha \in \{\sum, \prod, \min\}$ fulfills inclusion maximality.
\end{prop}

\begin{proof}
Let $\utilityFunction$ be a utility function satisfying super-set monotonicity and \alpharule a budgeting method maximizing either the sum, the product or the minimum over all the voters utilities. 
For any voter $\voterName_i$, and for any pair of feasible bundles $\bundle$ and $\bundle'$ such that $\bundle' \subset \bundle$, we call $\bundleIntersection{i}$ and $\bundleIntersection{i}'$ the common subsets between $\approvalSet{i}$ and $\bundle$ and $\approvalSet{i}$ and $\bundle'$ respectively. Since $\bundle' \subset \bundle$, we have $\bundleIntersection{i}' \subseteq \bundleIntersection{i}$.
Since both the sum, product and the minimum utility of the voters are non decreasing with the utility of individual voters, if $\bundle'$ is optimal for any of these rules, then $\bundle$ is also optimal. The budgeting method $\alpha\!-\!\mathcal{R}_u$ thus satisfies inclusion maximality.
\end{proof}

Note that when a budgeting method is resolute, meaning that it returns only one winning bundle, this axioms requires that the only winning bundle is exhaustive. This means that if we use tie-breaking mechanism to choose a solution among several optimal ones, they should select an exhaustive solution. Note that it can be easily obtained by adding projects greedily from an optimal solution that is not exhaustive.

The next two axioms focus on robustness, especially when projects have a composite structure (i.e. a large project can be divided into several small projects, or small projects merged into one large project). 

\medbreak

\begin{definition}
A budgeting method $\aggregationRule$ satisfies \emph{splitting monotonicity} if for every budgeting scenario $\budgetingScenario=(\projectSet,\voterSet,\costFunction,\budgetLimit)$, for each $\projectName_x \in  \aggregationRule(\budgetingScenario)$ and each budgeting scenario $\budgetingScenario'$ which is formed from $\budgetingScenario$ by splitting $\projectName_x$ into a set of projects $\projectSet'$ such that $\cost{\projectSet'}\!=\!\cost{\projectName_x}$, and such that the voters which approve $\projectName_x$ in $\budgetingScenario$ approve all items of $\projectSet'$ in $\budgetingScenario'$ and no other voters approve items of $\projectSet'$, it holds that $\aggregationRule(\budgetingScenario') \cap \projectSet' \neq \emptyset$.
\end{definition}

\medbreak

\begin{prop}
For $\alpha \in \{\sum, \prod, \min \}$, the budgeting method $\alpha - \aggregationRule_{u_M}$ fulfills splitting monotonicity.
\end{prop}
\begin{proof}
Let $\budgetingScenario=(\projectSet,\voterSet,\costFunction,\budgetLimit)$ be a budgeting scenario, and let 
$\bundle$ be the bundle returned by \alpharule\space for $\budgetingScenario$. Let  $\projectName_x$ be a project in the bundle \alpharule$(\budgetingScenario)$. Let $\budgetingScenario'\!=\!(\projectSet',\voterSet',\costFunction',\budgetLimit)$ be the budgeting scenario formed from $\budgetingScenario$ in which $\projectName_x$ is divided into a set $X'$ of projects such that $\cost{X'}\!=\!\cost{\projectName_x}$. Voters in $\voterSet'$ are the same than in $\voterSet$ except that any voter approving project $\projectName_x$ in $\voterSet$ approves all the projects of $X'$ in $\voterSet'$. The bundle 
$\bundle$ maximizes the objective of the rule \alpharule.  
Note that the utility of any subset that does not contain $\projectName_x$ is identical for $\budgetingScenario$ and $\budgetingScenario'$, and brings the same satisfaction to each voter: it therefore has the same quality regarding the aggregating criterion of \alpharule for $\budgetingScenario$ and $\budgetingScenario'$. 
Let $\bundle'$ be the bundle $\bundle$ in which $\projectName_x$ is replaced by all the projects in $X'$. Bundle $\bundle'$ is a feasible solution for $\budgetingScenario'$. Any voter $\voterName_i'$ in $\voterSet'$ has a corresponding voter $\voterName_i$ in $\voterSet$. We recall that $\bundleIntersection{i}$ denotes the set of projects that are common between a bundle $\bundle$ and the approval set of a voter~$\voterName_i$. There are two cases:\\
    $\bullet$ $ X' \cap \bundleIntersection{i}'\!=\!\emptyset$: in this case,  $\utilityMobius{\bundleIntersection{i}'}{\budgetingScenario'}=\utilityMobius{\bundleIntersection{i}}{\budgetingScenario}$ \\
    $\bullet$ $X' \subseteq \bundleIntersection{i}'$: we have $\cost{\bundleIntersection{i}'}=\!\cost{\bundleIntersection{i}}$ and $\rate{\bundleIntersection{i}'}{\voterSet}\!=\!\rate{\bundleIntersection{i}}{\voterSet}$. Additionally, we have $\prod_{b' \in \bundleIntersection{i}'} \rate{b'}{\voterSet'} \leq \prod_{b \in \bundleIntersection{i}} \rate{b}{\voterSet}$, since the rates do not change but the number of projects is larger 
     in $\bundleIntersection{i}'$ than in $\bundleIntersection{i}$. This is also true for any subset $C \subseteq \bundleIntersection{i}'$ such that $X' \subseteq C$. Therefore, because of the super-set monotonicity property, we have  $\utilityMobius{\bundleIntersection{i}'}{\budgetingScenario'} \geq \utilityMobius{\bundleIntersection{i}}{\budgetingScenario}$.\\
Overall, $\bundle'$ is at least as good as any solution containing no element of $X'$, meaning that either $\bundle'$ maximizes the rule criterion or a solution containing at least one project in $X'$ does. Therefore there is a $\projectName$ in $X'$ such that $\projectName$ is in \alpharule$(\budgetingScenario')$: the \alpharule\space rule fulfills splitting monotonicity.
\end{proof}

\medbreak

\begin{definition}
A budgeting method $\aggregationRule$ satisfies  \emph{merging monotonicity} if for each budgeting scenario $\budgetingScenario=(\projectSet,\voterSet,\costFunction,\budgetLimit)$, and for each $\projectSet' \subseteq  \aggregationRule(\budgetingScenario)$  such that for each $\voterName_i \in \voterSet$ we either have $\approvalSet{i} \cap \projectSet'\!=\!\emptyset$ or $\projectSet' \subseteq \approvalSet{i}$ -- i.e. a voter approves either all projects from $\projectSet'$ or none -- it holds that $\projectName \in \aggregationRule(\budgetingScenario')$ for $\budgetingScenario'\!=\!(\projectSet \setminus \{\projectSet'\} \cup \{\projectName\}, \voterSet',\costFunction',\budgetLimit)$, $\costFunction'(\projectName)\!=\!\costFunction(\projectSet')$,  and each voter $\voterName_i \in \voterSet$ for which $\projectSet' \subseteq \approvalSet{i}$ in $\budgetingScenario$ approves $\projectName$ in $\budgetingScenario'$, and no other voter approves~$\projectName$.
\end{definition}

\medbreak

\begin{prop}
Let $\alpha \in \{\sum, \prod, \min \}$. If a utility function $\utilityFunction$ fulfills the strong effect of positive synergy property and cost consistency, then the budgeting method \alpharule\space does not fulfill merging monotonicity.
\end{prop}
\begin{proof}
{$\bullet$ Case where  $\alpha\!=\!\Sigma$}. Let $T$ be an even positive integer. Let us consider a budgeting scenario $\budgetingScenario=(\projectSet,\voterSet,\costFunction,\budgetLimit)$ with $\projectSet\!=\!{x_1,x_2,y}$, $\cost{x_1}\!=\!\cost{x_2}\!=\!T/2$, $\cost{y}\!=\!T$ and $\budgetLimit\!=\!T$. There are two types of voters in $\voterSet$. There are $\nbVoters_1$ voters of the first type, and each one of them approves $x_1$ and $x_2$. There are $\nbVoters_2$ voters of type 2, and they all approve $y$ as shown in Figure~\ref{fig:merging_monotonicity_sum_1}. By cost consistency, we know that there exists a constant $k$ such that  $\utility{x_1}{\budgetingScenario}=\utility{x_2}{\budgetingScenario}=kT/2$ and $\utility{y}{\budgetingScenario}=kT$. By strong effect of positive synergies, we have $\utility{\{x_1,x_2\}}{\budgetingScenario}>\utility{\{x_1\}}{\budgetingScenario}+\utility{\{x_2\}}{\budgetingScenario}$ and consequently $\utility{\{x_1,x_2\}}{\budgetingScenario}>kT$. Let $\epsilon\!=\!\utility{\{x_1,x_2\}}{E}-kT>0$. The bundle $\{x_1,x_2\}$ has a total utility of $\nbVoters_1 (kT+\epsilon)$, the bundle $\{y\}$ has a utility of $\nbVoters_2kT$. If $\nbVoters_1 kT - \nbVoters_2 kT + \nbVoters_1\epsilon >0$, then $\{x_1,x_2\}$ is the best bundle.

\begin{figure}[H]
        \centering
        \begin{tikzpicture}
        \project{$x_1$}{3}{3}{1}
        \project{$x_2$}{3}{6}{1}

        \project{$y$}{6}{6}{0}

        \node[text width=1cm] at (-0.4,1.35) {$n_1$};
        \node[text width=1cm] at (-0.4,0.35) {$n_2$};

        \end{tikzpicture}
    \caption{First budgeting scenario $\budgetingScenario$}
    \label{fig:merging_monotonicity_sum_1}
\end{figure}
We now consider $\budgetingScenario'\!=\!(\projectSet',\voterSet',\costFunction',\budgetLimit)$ another budgeting scenario such that $\projectSet'\!=\!\{x,y\}$, $\costFunction'(x)\!=\!\cost{x_1}+\cost{x_2}\!=\!\costFunction'(y)\!=\!\cost{y}\!=\!T$. In $\voterSet'$ we create $\nbVoters_1$ voters approving $x$ and $\nbVoters_2$ voters approving $y$. Note that the budgeting scenario $\budgetingScenario'$ is similar to $\budgetingScenario$ except that the projects $x_1$ and $x_2$ have merged in a project of size $T$. Because of cost consistency, we have $\utility{\{x\}}{\budgetingScenario'}\!=\!\utility{\{y\}}{\budgetingScenario'}\!=\!kT$. Therefore the bundle $\{x\}$ has a total utility of $\nbVoters_1kT$ and the bundle $\{y\}$ still has a utility of $\nbVoters_2 kT$. If $\nbVoters_1 < \nbVoters_2$, $\{y\}$ is the winning bundle.

\begin{figure}[H]
        \centering
        \begin{tikzpicture}
        \project{$x$}{6}{6}{1}

        \project{$y$}{6}{6}{0}

        \node[text width=1cm] at (-0.4,1.35) {$n_1$};
        \node[text width=1cm] at (-0.4,0.35) {$n_2$};

        \end{tikzpicture}
    \caption{Second budgeting scenario $\budgetingScenario'$}
\end{figure}


By setting $\nbVoters_1\!=\!\lceil 2kT/\epsilon \rceil$ and $\nbVoters_2\!=\!\nbVoters_1+1$,  
$\{x_1,x_2\}$ is the winning bundle for $\budgetingScenario$ and $\{y\}$ is the winning bundle for $\budgetingScenario'$, giving us an instance for which the \sumrule\space rule does not fulfill merging monotonicity. 


\smallbreak
{$\bullet$ Case where  $\alpha\in\{\prod, \min\}$}.
Let us consider a budgeting scenario $\budgetingScenario=(\projectSet,\voterSet,\costFunction,\budgetLimit)$ with $\projectSet\!=\!\{x_1,x_2,x_3,x_4,y\}$, $\cost{x_1}\!=\!\cost{x_2}\!=\!\cost{x_3}\!=\!\cost{x_4}\!=\!(T-2)/4$, $\cost{y}\!=\!T/2+1$ with $T$ an even integer and $\budgetLimit\!=\!T$.  There are two voters in $\voterSet$: the first one approves of $x_1$, $x_2$, $x_3$ and $x_4$, the second one approves of $y$. 

\begin{figure}[H]
        \centering
        \begin{tikzpicture}
        \project{$x_1$}{1}{1}{1}
        \project{$x_2$}{1}{2}{1}
        \project{$x_3$}{1}{3}{1}
        \project{$x_4$}{1}{4}{1}

        \project{$y$}{2.5}{2.5}{0}

        \node[text width=1cm] at (-0.4,1.35) {$1$};
        \node[text width=1cm] at (-0.4,0.35) {$1$};

        \end{tikzpicture}
    \caption{First budgeting scenario $\budgetingScenario$}
    \label{fig:merging_monotonicity_prod_1}
\end{figure}

When maximizing either the $\min$ utility or the product, for any utility function $\utilityFunction$ fulfilling cost consistency and the strong effect of positive synergies, the winning bundle will be $y$ plus two projects $x_i$ and $x_j$. Let us assume, without loss of generality that the projects $x_1$ and $x_2$ are part of the winning bundle. Let $\budgetingScenario'\!=\!(\projectSet',\nbVoters',\costFunction',\budgetLimit)$ be a budgeting scenario formed from $\budgetingScenario$ in which projects $x_1$ and $x_2$ are merged into one project $X$ of cost (and therefore utility)  $T/2-1$. 

\begin{figure}[H]
        \centering
        \begin{tikzpicture}
        \project{$X$}{2}{2}{1}
        \project{$x_3$}{1}{3}{1}
        \project{$x_4$}{1}{4}{1}

        \project{$y$}{2.5}{2.5}{0}

        \node[text width=1cm] at (-0.4,1.35) {$1$};
        \node[text width=1cm] at (-0.4,0.35) {$1$};

        \end{tikzpicture}
    \caption{Second budgeting scenario $\budgetingScenario'$}
\end{figure}

The utilities of $x_3$ and $x_4$ are still $(T-2)/4$. By strong superadditivity of groups, the utility of $\{x_3,x_4\}$ is strictly larger than $2(T-2)/4$ and strictly larger than the utility of $X$ consequently. Therefore the winning bundle for $\budgetingScenario'$ is $\{y, x_3,x_4\}$. Since $X$ is not in this bundle, merging monotonicity is not fulfilled.
\end{proof}

The next axiom states that if the cost of a funded project decreases, it is still guaranteed to be funded. It is easy to see that this axiom is not compatible with the cost consistency property.

\medbreak

\begin{definition}
A budgeting method $\aggregationRule$  satisfies \emph{discount monotonicity} if for each budgeting scenario $\budgetingScenario=(\projectSet,\voterSet,\costFunction,\budgetLimit)$ and each item $b \in \aggregationRule(\budgetingScenario)$, it holds that $b \in \aggregationRule(\budgetingScenario')$ for $\budgetingScenario'\!=\!(\projectSet,\voterSet,\costFunction',\budgetLimit)$ where for each item $a \neq b$, $\costFunction'(a)\!=\!\cost{\projectName}$ and $\costFunction'(b)\!=\!\cost{b}-1$. 
\end{definition}

\medbreak

\begin{prop}
Let $\alpha \in \{\sum, \prod, \min \}$. If a utility function $\utilityFunction$ fulfills cost consistency, then the budgeting method \alpharule\xspace does  not fulfill discount monotonicity.
\end{prop}

\begin{proof}
Let us consider a budgeting scenario $\budgetingScenario=(\projectSet,\voterSet,\costFunction,\budgetLimit)$ with $\projectSet\!=\!{x_1,x_2,y}$, such that  $\cost{x_1}\!=\!4, \cost{x_2}\!=\!3$, $\cost{y}\!=\!4$ and $\budgetLimit\!=\!8$. There are two voters in $\voterSet$: the first one approves $x_1$ and $x_2$, and the second one approves $y$. When maximizing either the $\sum$, the $\min$ or the $\prod$ of utilities, for any utility function $\utilityFunction$ fulfilling cost consistency, the winning bundle will be $y$ plus project $x_1$. Let $\budgetingScenario'\!=\!(\projectSet,\voterSet,\costFunction',\budgetLimit)$ be a budgeting scenario formed from $\budgetingScenario$ in which project $x_1$ now has a cost of $2$ instead of $4$. The winning bundle is now $\{y,x_2\}$. The cost of project $x_1$ was reduced and it was removed from the winning bundle, therefore discount monotonicity is not fulfilled. 
\end{proof}

This last axiom states that any funded project in a winning bundle is still funded when the budget limit increases. 

\medbreak

\begin{definition}
A budgeting method $\aggregationRule$ fulfills \emph{limit monotonicity} if for each pair of budgeting scenarios $\budgetingScenario=(\projectSet,\voterSet,\costFunction,\budgetLimit)$ and $\budgetingScenario'\!=\!(\projectSet,\voterSet,\costFunction,\budgetLimit+1)$ with no item which costs exactly $l+1$, it holds that $\projectName \in \aggregationRule(\budgetingScenario) \implies \projectName \in \aggregationRule(\budgetingScenario')$.
\end{definition}

\medbreak

\begin{prop}
Let $\alpha \in \{\sum, \prod, \min \}$. 
If a utility function $\utilityFunction$ fulfills cost consistency, then the  budgeting method  \alpharule does not fulfill limit monotonicity.
\end{prop}

\begin{proof}
$\bullet$ Case where $\alpha\!=\!\sum$: Let us consider a budgeting scenario $\budgetingScenario=(\projectSet,\voterSet,\costFunction,\budgetLimit)$ with $\projectSet\!=\!{x_1,x_2,x_3}$, $\cost{x_1}\!=\!2, \cost{x_2}\!=\!5$, $\cost{x_3}\!=\!6$ and $\budgetLimit\!=\!6$. There are three voters in $\voterSet$: the first one approves of $x_1$, the second one approves of $x_2$ and the third one approves of $x_3$. When maximizing the sum of utilities, for any utility function $\utilityFunction$ fulfilling cost consistency, $\{x_3\}$ will be the winning bundle.
Let $\budgetingScenario'\!=\!(\projectSet,\voterSet,\costFunction,\budgetLimit')$ be a budgeting scenario formed from $\budgetingScenario$ but such that the budget limit $\budgetLimit'$ is now $7$ instead of $6$. The winning bundle is now $\{x_1,x_2\}$. The budget limit was increased and project $x_3$ was removed from the winning bundle, therefore limit monotonicity is not fulfilled.\\

\smallbreak
$\bullet$ Case where $\alpha \in \{\min, \prod\}$: Let us consider a budgeting scenario $\budgetingScenario=(\projectSet,\voterSet,\costFunction,\budgetLimit)$ with $\projectSet\!=\!{x_1,x_2,x_3}$, $\cost{x_1}\!=\!1, \cost{x_2}\!=\!2$, $\cost{x_3}\!=\!3$ and $\budgetLimit\!=\!4$. There are two voters in $\voterSet$: the first one approves of $x_1$ and $x_2$, the second one approves of $x_3$. When maximizing either the min or the product of utilities, for any utility function $\utilityFunction$ fulfilling cost consistency, $\{x_1,x_3\}$ will be the winning bundle.
Let $\budgetingScenario'\!=\!(\projectSet,\nbVoters,\costFunction,\budgetLimit')$ be a budgeting scenario formed from $\budgetingScenario$ but such that the budget limit $\budgetLimit'$ is now $5$ instead of $4$. The winning bundle is now $\{x_2,x_3\}$. The budget limit was increased and project $x_1$ was removed from the winning bundle, therefore limit monotonicity is not fulfilled.
\end{proof}

From Proposition~\ref{prop:utility_function_um} and propositions from Section~\ref{sec:pb_methods_axioms}, we get the following corollary.
\begin{corollary}
The rules \alpharule\xspace for $\alpha \in \{\sum,\min,\prod\}$  and $u\!=\!u_M$ fulfill \emph{inclusion maximality} and \emph{splitting monotonicity}. 
They do not fulfill \emph{merging monotonicity}, 
\emph{discount monotonicity} and \emph{limit monotonicity}.
\end{corollary}

\section{Complexity}

\label{sec:complexity}

We show in this section that, for each $\alpha \in \{\sum, \prod, \min \}$, problem \alphaproblem is NP-hard when there are synergies, and this even if all the projects have unitary cost and for a very general class of utility functions.  
This shows that synergies add complexity, since  problem \sumproblem\space is polynomially solvable when projects have the same cost and without synergies (i.e. when the function $\utilityFunction$ is linear). Indeed, without synergies and with unitary size projects, selecting the projects by decreasing number of votes maximizes the sum of the utilities of the voters. Let us now show that, with synergies, this problem is NP-hard even with very general utility functions. 
We start by proving a preliminary result for the \textsc{Clique} problem.

\medbreak

\begin{lem}
The \textsc{Clique} problem is strongly NP-complete even if it is restricted to graphs $G$ in which $ d_{\max}< \sqrt{m}$, where $m$ is the number of edges and $d_{\max}$ is the maximum degree of a vertex of $G$.
\label{lem:clique}
\end{lem}
\begin{proof}
The \textsc{Clique} problem is the following one. We are given an undirected graph $G\!=\!(V,E)$, with $V$ the set of $n$ vertices and $E$ the set of $m$ edges. We denote by $d_i$ the degree of a vertex $i$, and by $d_{\max}$ the maximum degree of any vertex in $V$. We are also given an integer $K$. The question is: does there exist a clique  of size $K$ in $G$?

This problem is known to be strongly  NP-complete~\citep{garey1979computers}, and we now show that it is still strongly NP-complete when the graph $G$ is such that $\sqrt{m} > d_{\max}$. 

We reduce the \textsc{Clique} problem in any graph into the \textsc{Clique} problem in a graph where $\sqrt{m} > d_{\max}$. Let $G$ and $K$ be an instance of the {\sc Clique} problem without any constraint on $m$ and $d_{\max}$. 
We first transform graph $G$ into a graph $G'$, as follows. Graph $G'$ is built from graph $G$ by ``copying" $G$ $m$ times, obtaining $m$ connected components: for any vertex $v_i$ in $V$, we create $m+1$ vertices $\{v_{i,0}, v_{i,1} \cdots v_{i,m}\}$ in $V'$, and for each edge $(v_i,v_j)$ in $E$, we create $m+1$ edges $\{(v_{i,0},v_{j,0}) \cdots (v_{i,m},v_{j,m}) \}$ in $E'$. 
We denote by $d^G_{\max}$ (resp. $d^{G'}_{\max}$) the maximum degree of a vertex of $G$ (resp. $G'$), and by $m$ (resp. $m'$) the number of edges in $G$ (resp. $G'$). We have $d^{G'}_{\max}\!=\!d^G_{\max}$ and $m'\!=\!(m+1)m$. Since $m\geq d^G_{\max}$ and $d^{G'}_{\max}\!=\!d^G_{\max}$, we have: $m'\!=\!m(m+1)\geq d^G_{\max}(d^G_{\max}+1)$. Therefore,  $\sqrt{m'} > d^{G'}_{\max}$. 
We now show that there is a clique of size $K$ in $G'$ if and only if there is a clique of size $K$ in $G$. 
\begin{itemize}
    \item Let us first assume that there is a clique $C\!=\!\{v_1, v_2 \cdots v_K\}$ of size $K$ in $G$. In that case, the set $C'\!=\!\{v_{1,0}, v_{2,0} \cdots v_{K,0}\}$ is clique of size $K$ in $G'$ since for any edge connecting two vertices $v_i$ and $v_j$ in $G$ we created an edge connecting $v_{i,0}$ and $v_{j,0}$ in $E'$. 
    
    \item Let us now assume that there exists a clique of size $K$ in  $G'$. Such a clique can only be formed by a set of vertices $\{v_{1,i},v_{2,i} \cdots v_{K,i}\}$ with a fixed $i$ since no edges in $E'$ connect two vertices $v_{k,i}$ and $v_{l,j}$ with $i \neq j$ by construction. If such a clique exists, then the subset $C\!=\!\{v_{1},v_{2} \cdots v_{K}\}$ in $G$ is a clique as well since if an edge $(v_{k,i},v_{l,i})$ exists in $E'$, an edge $(v_{k},v_{l})$ exists in $E$. Therefore $C$ is a clique of size $K$ in $G$ and the answer to the \textsc{Clique} problem is yes.
\end{itemize}
Since our problem is in NP, and that there exists a polynomial time reduction of the strongly NP-complete {\sc Clique} problem into the \textsc{Clique} problem when $\sqrt{m} > d_{\max}$, we conclude that the \textsc{Clique} problem is strongly NP-complete even when $\sqrt{m} > d_{\max}$. 
\end{proof}

\medbreak

\begin{prop}
Problem \sumproblem\space is strongly NP-hard, even if all the projects have unit costs. This is true if $\utilityFunction=\utilityFunctionMobius$, as well as for any utility function $\utilityFunction$  such that the utility of two projects  that have been selected together by at least one voter  is strictly larger than the utility of two projects approved by the same number of voters but that have never been selected together by a same voter.
\end{prop}

\nopagebreak[4]

\begin{proof}
The decision version of our problem, that we will denote \sumproblemdec, is the following one. We are given  a number $R \in \mathbbm{Z}$ and a budgeting scenario $\budgetingScenario\!=\!(\projectSet,\voterSet,\costFunction,\budgetLimit)$ with $\costFunction$ a cost function such that the cost of each project of $\projectSet$ is exactly one. We consider that the utility function $\utilityFunction$ is such that the utility of a pair of projects  selected at least once together is strictly larger than the utility of any other pair of projects that have been selected the same number of times but that have never been selected together. 
The set $\projectSet$ is a set of $\nbVoters$ voters $\{\voterName_1,\dots, \voterName_{\nbVoters}\}$, having each one approved up to $\budgetLimit$ projects of $\projectSet$. 
The  question is: does there exist a set $B \subset \projectSet$ of up to $\budgetLimit$ projects such that the utility of $\bundle$,  $\sum_{\voterName_i \in \voterSet} \utility{\bundleIntersection{i}}{\budgetingScenario}$, is at least $R$ ?  

\smallskip

We reduce the strongly NP-complete problem {\sc Clique} to this problem. We will assume that the instance of the \textsc{Clique} problem is a graph such that $\sqrt{m} > d_{\max}$ (the \textsc{Clique} problem is still NP-complete in this case, as shown by Lemma \ref{lem:clique}). The {\sc Clique} problem is as follows: given an graph $G\!=\!(V,E)$, such that $\sqrt{m}>d_{\max}$, and an integer $K$, the question is: does there exist a clique of size $K$?

Given an instance ($G,K$) of the {\sc Clique} problem, we create an instance of \sumproblemdec as follows. 
We first transform graph $G$ into a graph $G'$, as follows. We start by setting $G'\!=\!G$, and we assume that the $|V|$ vertices of $G'$ are labelled $\{1,\dots,|V|\}$. For each vertex $i$ of degree $d_i<d_{\max}$, we add $(d_{\max}-d_i)$ new neighbor vertices, denoted by ${Dummy}({i,1}), \dots Dummy({i,d_{\max}-d_i})$. By doing this, the vertices of $\{1,\dots,|V|\}$ are all of degree $d_{\max}$.  Let $G'\!=\!(V',E')$ be the graph obtained. Each newly added vertex $Dummy({i,j})$ is of degree 1 in $G'$. Therefore, the number of newly added vertices in $G'$ is $n_{dummy}\!=\!\sum_{i\!=\!1}^{|V|} d_{\max} - d_i\!=\!|V|d_{\max}-2|E|$, and the number of newly added edges is the same value. 
We label the newly added  vertices (if any) as $\{|V|+1, \dots , |V| + n_{dummy}\}$. 



We now create from $G'$ a set of projects $\projectSet$ as follows.  To each vertex $i\in\{1,\dots,|V'|\}$ we create a corresponding project $P_i$ of cost 1: there are thus $|V|$ projects corresponding each one to one vertex of $V$, and $n_{dummy}$ projects corresponding each one to one dummy vertex. 
We create a set $\mathcal{V}$ of $m_{{V}}\!=\!|E'|+ (d_{\max}-1)n_{dummy}$ voters. To each edge $\{x,y\}\in E'$, we create a voter which approves exactly two projects: projects $P_x$ and $P_y$, corresponding to vertices $x$ and $y$. For each dummy vertex, we create $(d_{\max}-1)$ voters that approve only the project corresponding to the dummy vertex. 

We fix the maximum budget to $l\!=\!K$ (since all the projects have a unitary cost, this means that up to $K$ projects can be selected).  
The value of $R$, the target utility, depends of the synergy function. 
We observe that in our instance of \sumproblemdec each project is chosen by the same number of voters ($d_{\max}-1$). 
Let $u_{together}$ be the utility that a voter obtains for a set of two projects which have both been chosen by the voter. The sequel of the proof works for all utility function such that $u_{together}> 2$. This is in particular true for $\utilityFunctionMobius$, as shown by the following fact.  
\smallskip

{\bf Fact 1: } If the utility function is $\utilityFunctionMobius$, then $u_{together}> 2$.\\
\emph{Proof of the fact: }  
Let us show that the utility function $\utilityFunctionMobius$ count positive interactions for pairs of projects corresponding to vertices connected by an edge in $G'$. 
For function $\utilityFunctionMobius$, we have:
$$m(\{x,y\},\mathcal{V})\!\geq\!r(\{x,y\},\mathcal{V})-r(\{x\},\mathcal{V})r(\{y\},\mathcal{\mathcal{V}})
$$
$$m(\{x,y\},\mathcal{V})\!\geq\!\frac{1}{m_{V}}-\frac{(d_{\max})^2}{(m_V)^2}=\frac{1}{m_V}\left(1-\frac{(d_{\max})^2}{m_V}\right)$$

Furthermore: 
$$m_{{V}}\!=\!|E|\!+\!\sum_{i\!=\!1}^{|V|} d_{\max}(d_{\max}-d_i)\!=\!|E|\!+\!|V|  (d_{\max})^2\!-\!\sum_{i\!=\!1}^{|V|} d_{\max}  d_i$$


Since $d_id_{\max}\!\leq\!(d_{\max})^2$, we get $\sum_{i\!=\!1}^{|V|} d_id_{\max}\!\leq\!|V|(d_{max})^2$, and thus
$m_V\geq |E|$. 
Since $\sqrt{|E|} > d_{max}$, $m_V>(d_{\max})^2$ and the Möbius transform of the pair is larger than $0$, meaning that the utility of the subset $\{x,y\}$ is larger than the sum of their costs: $u_{together}>2$. 
\smallskip

Let us now show that it possible to select a set of at most $K$ projects of total utility larger than or equal to $R\!=\! K d_{max} + \frac{K(K-1)}{2} (u_{together}-2)$ if and only if there is a clique of size $K$ in $G$.  
    $\bullet$ Let us first assume that there is a clique $C$ of size $K$ in $G$. Let $S^{clique}$ be the set of the $K$ projects which correspond to the $K$ vertices of $C$. 
    Note that for each couple of projects $x$ and $y$ of $S^{clique}$, exactly one voter has approved both $x$ and $y$. The utility of $S^{clique}$ is thus $u_{together}$ for each of these $K(K-1)/2$ voters. Note also that each project has been selected by exactly $d_{max}$ voters. Therefore, for the $K d_{max} - 2 \times K(K-1)/2$ voters who approve exactly one project of $S^{clique}$, the utility of $S^{clique}$ is 1. The other voters do not approve any project of $S^{clique}$ and have a utility of 0. The total utility of $S^{clique}$ is thus $1\times (K d_{max} - 2 \times \frac{K(K-1)}{2}) + u_{together}\frac{K(K-1)}{2}\!=\! K d_{max} + \frac{K(K-1)}{2} (u_{together}-2)\!=\!R$. The answer to our problem is thus `yes'.
    
    
    $\bullet$ Let us now assume that there is a set $C$ of at most $K$ projects of  total utility  at least $R\!=\!K d_{max} + \frac{K(K-1)}{2} (u_{together}-2)$. 
    Note that each project is approved by exactly $d_{max}$ voters. The utility of  $C$ for a given voter is 0 if the voter does not select any project of $C$, 1 if it selects exactly one project, and $u_{together}>2$ if it approves exactly two projects (recall that a voter approves at most 2 projects). The utility of $C$ is thus equal to $n_1$, the number of voters who approve exactly one project of $C$, plus $n_2 \times  u_{together}$, where $n_2$ is the number of voters who approve exactly two projects of $C$. We have $R\!=\!K d_{max} - 2 \frac{K(K-1)}{2} + \frac{K(K-1)}{2} u_{together}\leq n_1+ n_2 u_{together}$, and $n_1+2n_2 \leq K d_{max} $ (since $n_1+ 2 n_2$ is equal to the total number of votes for projects of $C$ and $C$ is of size at most $K$). Therefore, $n_1\!=\!K d_{max} - 2 \frac{K(K-1)}{2}$, and $n_2\!=\!\frac{K(K-1)}{2}$. This means that there are exactly $K$ projects in $C$ and that for each couple of projects of $C$, there is a voter who approves both projects (recall that there is exactly one voter by edge in $G'$). Therefore, there exists a clique of size $K$ in $G'$, and thus a clique of size $K$ in $G$.  
There exists a polynomial time reduction of the strongly NP-complete {\sc Clique} problem into the decision version of our problem: \sumproblemdec is thus  strongly NP-hard. 
\end{proof}


The next result extends the result from 
\cite{sreedurga2022maxmin}, 
which proves that the maxmin participatory budgeting problem is strongly NP-hard for approval voting when the utility function is the sum of the costs of the funded approved projects. We generalize this result by proving that this is true for both the maxmin and the product of utilities and we show that we only need a very weak condition on the utility function for this to be true. Additionally, it holds for knapsack voting, which is more specific that approval voting. We also show that the problem is hard to approximate. We first prove the following lemma.

\medbreak

\begin{lem}
The \textsc{Set cover} problem is strongly NP-complete even when restricted to instances in which the number of subsets containing the same element is bounded by $K$, the size of a feasible solution.
\label{lem:set_cover}
\end{lem}

\begin{proof}
The \textsc{Set cover} problem is the following one: we are given a set $\mathcal{U}$ of $n$ elements, called the universe, and a collection $S$ of $m$ sets whose union is $\mathcal{U}$. Given an integer $K<m$, the question is: does there exist a set $\mathcal{S}$ of elements in $S$, such that $\cup_{s \in \mathcal{S}}\!=\!\mathcal{U}$ and $|\mathcal{S}| \leq K$ ? 

From an instance $\mathcal{U}$,$S$,$K$, we create a new instance $\mathcal{U}'$,$S'$,$K'$. In this new instance, we create $m$ dummy elements $\{x^1_{dummy} \cdots x^m_{dummy}\}$ and $m$ dummy sets $\{s^1_{dummy} \cdots s^m_{dummy}\}$ such that $s^i_{dummy}$ contains $x^i_{dummy}$. We then have $n'\!=\!n+m$ and $\mathcal{U'}\!=\!\mathcal{U} \cup \{x^1_{dummy},  \cdots, $ $ x^m_{dummy}\}$, $m'\!=\!2m$ and $S'\!=\!S \cup \{s^1_{dummy} \cdots s^m_{dummy}\}$ and $K'\!=\!K+m$. In this new instance, it is easy to see that each element is contained by at most $m<K'$ sets. 

We now prove that there exists a set cover of $\mathcal{U}'$ with subsets of $S'$ and of size $K'$ at most if and only if there exists a set cover of $\mathcal{U}$ with subsets of $S$ and of size $K$ at most.

\begin{itemize}
    \item Let us first suppose that there exist a cover $C'$ of $\mathcal{U}'$ with subsets of $S'$ and of size $K'$ at most. This cover necessarily contains the $m$ dummy sets since these sets are the only one containing the $m$ dummy vertex. The $K'-m<K$ other sets of the cover form a feasible cover of the $n$ elements of $\mathcal{U}$ and all of these sets are in $S$.
    \item Now, we suppose that that there exist a cover $C$ of $\mathcal{U}$ with subsets of $S$ and of size $K$ at most. The elements of $\mathcal{U}'$ that are not covered by $C$ are the dummy elements. By adding the $m$ dummy sets of $S'$ to $C$, we obtain a cover $C'$ covering all the elements from $\mathcal{U}$ plus the $m$ dummy elements and of size of $|C|+m$. Since $|C| \leq K$, we have $|C|+m \leq K+m\!=\!K'$, we therefore have a feasible cover of $\mathcal{U}'$.
\end{itemize}
There exist a polynomial time reduction between any instance of \textsc{Set cover} to a version of the \textsc{Set cover problem} in which the number of sets containing the same element is bounded by $K$. Therefore the \textsc{Set cover} is still strongly NP-complete in that case.
\end{proof}

\medbreak

\begin{prop}
Problems \minproblem\space and \prodproblem\space are strongly NP-hard for any utility function $\utilityFunction$ such that $\utility{\emptyset}{E}\!=\!0$ and $\utility{S}{\budgetingScenario}>0$ for each $S \neq \emptyset$. For any $\delta>1$, there is no polynomial time $\delta$-approximate algorithm if $P\neq NP$.
\end{prop}

\begin{proof}

The decision version of our problem is the following one. We are given a real number $R$ and a budgeting scenario $\budgetingScenario=(\projectSet,\voterSet,\costFunction,\budgetLimit)$ with $\projectSet$ a set of $\nbProjects$ projects and $\costFunction$ a cost function such that the cost of each project is exactly one. We consider a utility function $\utilityFunction$ such that $\utility{S}{\budgetingScenario}>0$ if $S \neq \emptyset$.
The set $\voterSet$ is a set of $\nbVoters$ voters $\{\voterName_1,\dots, \voterName_{\nbVoters}\}$, having each one approved up to $\budgetLimit$ projects of $\projectSet$. The  question is: does there exist a set $B \subset \projectSet$ of up to $\budgetLimit$ projects such that the utility of $\bundle$,  $\prod_{\voterName_i \in \voterSet} \utility{\bundleIntersection{i}}{\budgetingScenario}$ (or $\min_{v \in V} \utility{\bundleIntersection{i}}{\budgetingScenario}$), is at least $R$ ?

We will reduce the strongly NP-complete problem {\sc Set cover}~\citep{garey1979computers} to this problem. The \textsc{Set cover} problem is the following one: we are given a set $\mathcal{U}$ of $n$  elements, called the universe, and a collection $S$ of $m$ sets whose union equals the universe. Given an integer $K$, the question is: does there exist a set $\mathcal{S}$ of sets in $S$, such that $\cup_{s \in \mathcal{S}}\!=\!\mathcal{U}$ and $|\mathcal{S}| \leq K$ ? 
We suppose that the number of subsets containing the same element is bounded by $K$ -- as shown by Lemma~\ref{lem:set_cover}, the problem is still strongly NP-complete in that case.

Let $\mathcal{U}$, $S$ and $K$ be an instance of the {\sc Set cover} problem. Let us create an instance of our problem. %

For each element $s$ in $\mathcal{U}$, we create a voter $\voterName_e$. For every subset $s$ in $S$, we create a project $\projectName_s$ of cost 1. This project is approved by any voter $\voterName_e$ such that $e \in s$. Note that, since the number of sets containing the same element is smaller than or equal to $K$, the number of projects approved by a voter is smaller than or equal to $K$. We set $\budgetLimit\!=\!K$ and $R\!=\!\epsilon$ with $\epsilon>0$. The question is now: does there exist a bundle $\bundle$ of projects such that the product (or minimum) of the voters' utilities for bundle $\bundle$ is greater than or equal to $\epsilon$ ? Since $\epsilon$ can be as small as we want, we can simply look for a solution with value strictly larger than $0$.\\

We show that there is a positive answer to this question if and only if there exists a cover  of size $K$ in $S$. 
\begin{itemize}
    \item Let us first assume that there is a cover $C$ of size $K$ in $S$. Let $B^{cover}$ be the set of the $K$ projects which correspond to the $K$ sets of $S$. All voters have at least one of their approved projects in the bundle $B^{cover}$, since the projects corresponding to the sets have been chosen by the voters matching with the elements. Therefore, if a voter did not have at least one approved project in $B^{cover}$, then the cover $C$ would not cover the element corresponding to the voter. The answer to our problem is thus `yes'.
    \item Let us now assume that it possible to select at most $K$ projects such that the total utility is at least $R\!=\!\epsilon$. Since we use the product or the min, this means that every voter has at least one of her approved projects in the funded bundle $\bundle$. We know that for each $\voterName_e \in \voterSet$, there is one project of $\approvalSet{e}$ in $\bundle$. If we consider the cover $C^B$ formed by the sets corresponding to the projects in $\bundle$, this means that for every element $e$, there is a subset $s$ in $C^B$ such that $e \in s$. Since the size of $\bundle$ is at most $K$, the size of $C^B$ is at most $K$, which means that $C^B$ is a feasible cover for the \textsc{Set cover} problem. The answer is thus `yes'. 
\end{itemize}
There exists a polynomial time reduction of the strongly NP-complete {\sc Set cover} problem into our problem: our problem is strongly NP-hard.  
Furthermore, a $\delta$-approximate algorithm, with $\delta$, would allow to detect whether there exist a solution with a product (or minimum) of utilities strictly larger than $0$, and thus would allow to solve the \textsc{Set cover} problem. 
Therefore, for any $\delta>0$, there does not exist polynomial time $\delta$-approximate solution for our problem, unless $P\!=\!NP$.  
\end{proof}

\section{A branch and bound algorithm}

In this section, we propose an exact branch and bound algorithm for \alpharule for $\alpha \in \{\sum,\prod,\min\}$ since, as shown in the previous section, this is NP-hard. We also run experiments on real-life instances.

\label{sec:branch_and_bound}
\subsection{Description of the algorithm}
Let us now present a branch and bound algorithm which solves \alphaproblem  exactly, for $\alpha \in \{\sum,\min,\prod\}$. 
Each level of the decision tree corresponds to a project:  we either add it to the funded projects -- if it fits in the remaining budget, 
or we ban it for the current node and all of its sons. In such a decision tree, each leaf corresponds to a feasible bundle. 
Since every decision is binary and there are $\nbProjects$ consecutive decisions, corresponding to the $\nbProjects$ projects, there are $2^{\nbProjects}$ leaves corresponding to the $2^{\nbProjects}$ possible subsets. Since the cost of an optimal bundle is at most $\budgetLimit$, at a current node, we add a project only if its cost is at most $\budgetLimit$ minus the cost of the currently funded projects -- this allows us to prune the tree. Moreover, at each node, we compute a feasible solution, and an upper bound of the value of the quality (w.r.t. the objective function of \alphaproblem) of a bundle that is reachable from this  node. If the upper bound of the value of a reachable bundle is smaller than the value  of a feasible solution we already know, then exploring the node's sons is useless, and we prune the tree. 

\smallskip

\noindent \textbf{Case where $\alpha=\sum$.} 
We compute a new feasible solution using a greedy rule,     called $\mathcal{R}^g_{|B_v|}$ by~\cite{talmon2019framework}, 
and which simply selects the projects by decreasing number of selections. At each node we consider the non yet considered projects by decreasing number of selections, and we add a project if it fits in the remaining budget. As we will see in Section~\ref{sec:expe}, using this algorithm at the root of the tree can also be used as a good and fast heuristic. 

The upper bound follows the same principle than the classic upper bound for the \textsc{Knapsack} problem, it is a linear relaxation.
In order to compute our upper bound, we need an upper bound on the utility that each project can give to a voter. By multiplying it by the number of voters who selected this project, we obtain an upper bound of the utility that a project can bring to the whole set of voters.

Before starting the exploration of the decision tree, for each project $\projectName$, we compute the sum of the Möbius transforms of each feasible subset in which $\projectName$ appears, divided by the size of this subset. This is an upper bound of how much utility a project can provide to one voter, we multiply it by the number of voters who selected this project, and obtain an upper bound of how much utility the project can bring to the whole set of voters. 
Note that this can be applied to other utility functions since the Möbius transforms can be computed for any utility function.

At each node, we then run the greedy algorithm selecting the (non yet selected nor eliminated) projects by decreasing upper bounds and we relax the integrity constraint, obtaining a fractional solution. This gives us an upper bound of the best solution that can be obtained at the current node. 
Note that the $k$-additivity assumption is particularly useful here since the maximum utility a project can give decreases when $k$ decreases, since all the Möbius transforms of subsets of size strictly greater than $k$ are null. 

\smallskip
\noindent \textbf{Case where $\alpha\in\{\min,\prod\}$.}  
We compute a feasible solution as follows: we look for the set of least satisfied voters. We choose the most frequently selected project by these voters, among projects that fits into the remaining budget. We repeat this process until there is no budget left. 

For the upper bound: at the root of the decision tree, we assume that each voter gets her favorite set of projects. At each node, 
we consider that each voter gets the projects that she voted for among the already selected projects, plus all the  projects that she selected among projects that still fits in the budget and which have not been considered yet. 
For example, if the selected projects cost half the budget, then any project costing more than half the budget could not be chosen and is therefore banned. If a project is banned, then we simply add it to the ban list. Then, we remove all newly banned project from the best reachable subsets of the voters.
This gives us an upper bound of the value of any reachable solution.

\smallskip
\noindent\textbf{Computing the utilities.} The utility provided by a given solution $\bundle$ to a voter $\voterName_i$ is the utility of $\bundleIntersection{i}$. Determining $\bundleIntersection{i}$ and computing its utility can be done in polynomial time if we know the utility function. Therefore, for each node of the decision tree, computing solutions and determining their value as upper and lower bounds can be done in polynomial time. 

To determine the utility of a bundle with function $\utilityFunctionMobius$, we use Equation~\ref{eq:utility_mobius}. Because of its recursive nature, we compute first, as a preprocessing step, the utility of singletons, then pairs, then triplets and so forth. Determining the utilities in this way cost up to $2^{\nbProjects}$ (since there are $2^{\nbProjects}$ subsets) times $\nbProjects\nbVoters$ operations (since determining the appearance rate of a subset costs $\nbProjects\nbVoters$ operations). This calculation is much faster with the $k$-additivity hypothesis, stating, as seen earlier, that we can consider interactions only in subsets of projects of size at most $k$.

With the $k$-additivity hypothesis, it is possible to know the utility of a subset of size $j$ in $O(j^k)$ operations, since its utility is the sum of all the Möbius transforms of its parts, and there are at most $j^k$ parts with a non null Möbius transform. This hypothesis has great implications on the computational side. 

\subsection{Experiments}
\label{sec:expe}

We use real instances from the Pabulib \citep{stolicki2020pabulib} library with a budget limit on the approbation sets of the voters. Experiments are run on an Intel Core i5-8250U processor with 8GB of RAM. We study the completion time of our algorithm and the impact of the synergies on the returned solutions.  We consider that $\alpha=\sum$ for the experiments since the sum is the most common aggregator.

\paragraph{Quality of the heuristic.} 
On average, the solution returned by the exact (branch and bound) algorithm has an overall utility 0.28\% higher than the utility of the solution returned by the heuristic $\mathcal{R}^g_{|B_v|}$ for the $\utilityFunctionMobius$ function: the heuristic returns, on the instances of Pabulib, very good solutions with regards to our optimization criterion.
\paragraph{Impact of the $k$-additivity assumption.} 
The $k$-additivity assumption allows to decrease the calculation time significantly -- the lower $k$ is, the fastest is the algorithm. Table~\ref{tab:times_k_additive} indicates the computation times obtained when $k\!=\!1$ (no synergy), and when $k\!=\!2$ and $k\!=\!3$ with utility function $\utilityFunctionMobius$. 

\begin{table}
    \centering
    \begin{tabular}{cccccc}
              Function & $\nbProjects\!=\!5$ & $\nbProjects\!=\!8$ & $\nbProjects\!=\!10$ & $\nbProjects\!=\!12$ & $\nbProjects\!=\!15$\\
        \hline
        $1$-additive & 0.013 & 0.057 & 0.10 & 0.26 & 0.77\\
        $2$-additive & 0.015 & 0.076 & 0.15 & 0.60 & 3.60\\
        $3$-additive & 0.016 & 0.11 & 0.25 & 1.43 & 9.85\\
    \end{tabular}
    \caption{Completion time (s) of the branch and bound algorithm.}
    \label{tab:times_k_additive}
\end{table}

\paragraph{Impact  of considering synergies.} We compare the optimal solution for the overlap utility function ($1$ additive) and the $\utilityFunctionMobius$ function with no $k$-additivity assumption. The optimal solutions are different in 35\% of the instances,  and the amount of money spent differently on average for all the instances is of 28.5\%. Therefore, taking synergies into account impacts the returned bundle in a little bit more than a third of the instances, and this impact may be important since the returned bundle considering synergies then differs significantly from an optimal bundle ignoring synergies.   

\section{Conclusion and future works}
This paper represents a first step towards taking project interactions into account in participatory budgeting problems. 
We introduced a utility function $\utilityFunctionMobius$ based on the frequency of selection of groups of projects by the voters, and we showed that it fulfills desirable axioms. We furthermore showed that taking into account synergies is NP-hard with the main aggregation criterion, and this for very general utility functions. We designed an exact algorithm that we implemented with $\utilityFunctionMobius$ but which can also be used with others utility functions.


Whereas, for very costly projects, decision makers will probably identify synergies ``by hand'', when there are numerous small projects, the authorities will likely be unable or unwilling to identify the synergies.  In such settings, identifying the synergies thanks to the preferences of the voters, is promising. 
We could also imagine settings where a community decides to use a participatory budgeting approach to set a program of a maximum fixed total duration $\budgetLimit$ among various events (presentations, courses, documentaries, etc), each event having a duration (considered as a cost). Members of the community could be  asked to select the events they prefer, using knapsack voting:  this situation is a participatory budgeting problem for which it would be particularly interesting to take into account synergies between the events. 


There are numerous future work directions. For example, it would be useful to design utility functions that fit as much as possible to the reality experienced by the users. Another challenging direction would be to  design algorithms that take into account synergies while ensuring proportional representation of groups of voters.

\bibliography{sn-bibliography}

\end{document}